# Influence of design parameters of upstream Venturi pipeline on multiphase flow measurement

Mengke ZHAN[a,b], Muhammad Alif bin RAZALI[b], Ayush MOITRA[a,b], Cheng-Gang XIE[b], Wai Lam LOH[c] and Jian-Jun SHU[a,*]

[a] School of Mechanical & Aerospace Engineering, Nanyang Technological University, 50 Nanyang Avenue, Singapore 639798;

[b] Schlumberger Oilfield (Singapore) Pte Ltd, Singapore Well Testing Center, 1 Benoi Crescent, Singapore 629986;

[c] Department of Mechanical Engineering, National University of Singapore, 9 Engineering Drive 1, Singapore 117575

## ABSTRACT

A multiphase flowmeter (MPFM) is used in the upstream oil and gas industry for continuous, in-line, real-time, oil-gas-water flow measurement without fluid separation. An MPFM typically consists of phase-fraction (holdup) and velocity (or flow rate) measurements. It is desirable to have homogeneous flow at the measurement location so that the phase-fraction measurement is representative. A horizontal blind-tee pipe-section is often installed to homogenize flow in the downstream vertical Venturi-based flowmeters; however, little information is available on the effect of horizontal blind-tee depth (HBD) on flow homogeneity. In addition, the Venturi vertical entrance length (VEL) leading to the Venturi inlet from the horizontal blind-tee outlet is another design parameter that may potentially affect the downstream phase distribution. The phase-fraction measurement principle requires liquid properties (*e.g.* water salinity). The local liquid richness makes the horizontal blind-tee an ideal location for measuring liquid properties; however, an excessive HBD may affect the reliability of the measurements of liquid properties, because local vortices may degrade liquid measurement representativeness if the local liquid velocity is too low. This study uses a computational fluid dynamics approach to evaluate the effect of HBD and VEL on multiphase flow measurement, including the Venturi differential-pressure, the Venturi inlet and the throat phase-fraction, and the local liquid-property at the end of a horizontal blind-tee. The computational results are validated with experimental data collected in a multiphase flow facility. Appropriate HBD and VEL are recommended.

---

*  Corresponding author.
   *E-mail address:* mjjshu@ntu.edu.sg (Jian-Jun SHU).

## 1  Introduction

A multiphase flowmeter (MPFM) is used in the upstream oil and gas industry to perform continuous, in-line, real-time, oil-gas-water flow rate measurement without fluid separation (bin Razali *et al*., 2021). An MPFM typically consists of phase-fraction (holdup) and velocity (or flow rate) measurements. Some phase-fraction measurement techniques include gamma-ray phase-fraction measurements that exploit the different mass attenuation properties of oil, water and gas, and electromagnetic techniques that measure the permittivity and conductivity of multiphase flow mixtures. Velocity can be measured by a differential-pressure device, the ultrasonic Doppler, or cross-correlation techniques (Huang *et al*., 2013). A Venturi flowmeter is a preferable differential-pressure device with less erosion concern compared to other variable cross-sectional area flow measurement devices, such as nozzles, orifices, and V-cone flowmeters. In addition, there are no moving or intrusive components in the Venturi flowmeter. This means that the Venturi flowmeters are not as heavily reliant on distribution and flow stability as the ultrasonic Doppler or cross-correlation-based measurements (Huang *et al*., 2013).

Homogeneous flow, where individual phases are well mixed, or flows with axial symmetry are desirable for accurate phase-fraction measurements at minimal manufacturing and environmental cost, as it allows for phase-fraction measurements (*e.g.* a single gamma-ray chordal beam) along any axis of symmetry. A horizontal blind-tee is a common way to homogenize gas-liquid flow when multiphase flow transitions from horizontal section to vertical upward flow in a vertically mounted MPFM (Hjertaker *et al*., 2018). In addition, the lower region of the horizontal blind-tee is mostly liquid-rich, making it an ideal location for liquid-property measurements (Pinguet *et al*., 2014; Fiore *et al*., 2019; Han *et al*., 2020). Measurements of liquid properties such as salinity can enhance MPFM measurements by monitoring changes in produced water in real-time.

Design parameters were investigated, such as the degree of mixing as a function of vertical position in vertical straight pipes (Zeghloul *et al*., 2015; Hjertaker *et al*., 2018). It was found that as the vertical length downstream of the horizontal blind-tee increases, the flow becomes increasingly axisymmetric (Hjertaker *et al*., 2018) and tends to regain its intermittency after a perturbation of 7 to 20 pipe-diameters (Zeghloul *et al*., 2015). It has been reported that flow velocity also affects phase distribution, with axisymmetric flow developing at shorter distances downstream of the horizontal blind-tee, with lower rather than higher flow velocities (Hjertaker *et al*., 2018); however, no studies have been done to characterize the effect of the Venturi vertical entrance length (VEL) on the phase-fraction measurement in the Venturi throat section, where mixing is better due to local flow velocities higher than in straight pipe-sections, such as the Venturi inlet. Horizontal blind-tee depth (HBD) is another design parameter that can affect local flow velocity and fluid exchange. For example, there may not be sufficient fluid exchange due to low local flow rate with excessive HBD, which may limit downstream flow mixing. Conversely, if the HBD is too short, the liquid and gas may still mix well, which may adversely affect measurements of liquid properties that must be performed in liquid-rich regions. The effect of HBD on the homogeneity effect of the horizontal blind-tee has only been studied in single-phase flow (Han *et al*., 2020; 2022). There are no studies to characterize the appropriate HBD required for multiphase measurements.

The motivation of this study is to address a research gap in the horizontal blind-tee design for a vertically mounted Venturi-based MPFM. This study aims to identify the effect of variation in (i) VEL and (ii) HBD on flow homogenization, phase-fraction and differential-pressure measurement in a Venturi, and (iii) the effect of variation in HBD on local liquid flow velocity and phase-fraction that may affect the measurements of liquid properties.

## 2    Methodology

### 2.1    Computational fluid dynamics model

Many computational fluid dynamics (CFD) studies are conducted on gas-liquid flow. Some of the most common modeling approaches include the Eulerian–Eulerian model (Shu & Wilks, 1995; Yamoah *et al.*, 2015; Zhang *et al.*, 2019; Acharya & Casimiro, 2020), the volume of fluid model (Shu, 2003b; Laleh *et al.*, 2011; López *et al.*, 2016) and the mixture model (Shu *et al.*, 1997; Shu, 2003a; Shang *et al.*, 2015). In this study, the Eulerian–Eulerian model was used as a modeling approach to solve the ensemble-averaged mass and momentum transport equations for dispersed gases, as well as continuous liquids, allowing detailed modeling of phase interactions. Drag (Tomiyama *et al.*, 1998), lift (Tomiyama *et al.*, 2002), wall-lubrication force (Frank *et al.*, 2008) and turbulent dispersion force (Burns *et al.*, 2004) were all included in the phase interaction forces. A mixture shear stress transport turbulence (SST $k-\omega$) model was used in the study (Menter, 1994; Cokljat *et al.*, 2006). The study found that the standard $k-\omega$ model is sensitive to the values of $k$ and $\omega$ outside the shear layer (Wilcox, 1997). Hence, an SST $k-\omega$ model (Menter, 1994) activating the $k-\omega$ model in the near-wall region and the $k-\varepsilon$ model far from the wall was used in the study. Apart from having a better accuracy outside the shear layer, the SST $k-\omega$ model also has a modified definition of the turbulent viscosity to account for the transport of turbulent shear stress. This improves the prediction of turbulent flow in the Venturi throat compared to the standard $k-\omega$ model, because the shear stress in the throat section (straight pipe) with low local strain rate, is due to the shear stress transport from the Venturi convergence section. Neglecting the transport of the turbulent shear stress would cause an underestimation of local shear stress and local turbulence levels. In addition, studies (Coughtrie *et al.*, 2013) have shown that the SST $k-\omega$ model has more accurate predictions of velocity, flow separation, and reattachment that also exist in the Venturi flow than other turbulence models such as the renormalization group $k-\varepsilon$ model. While some other turbulence models such as the Reynolds stress equation model also provide accurate predictions of shear stress transport, flow separation, and reattachment, the dual equations in the SST $k-\omega$ model are a more computationally efficient choice of turbulence models. The simulation is performed in three-dimensional geometry, as shown in Figure 1. Flow enters the horizontal section from the bottom left, through the horizontal-to-vertical blind-tee and exits the horizontal section from the top right. VEL is defined as the distance between the centerline of the horizontal pipe and the start of the convergent section of the Venturi. HBD is defined as the centerline of the vertical pipe to the end flange of the horizontal blind-tee. Note that the vertical Venturi consists of the Venturi inlet, leading to the Venturi convergence section and the Venturi throat before the two Venturi divergence sections. The use of the second divergence section is to reduce the materials needed for the long, smooth single divergence section leading to the outlet pipe. Pipe diameter $D$ is 2 inches. The use of a second vertical blind-tee downstream of the Venturi is to provide an access port to enable in-situ fluidic calibration (by inserting a calibration tool) of the gamma-ray Venturi MPFM. Having a second vertical blind-tee downstream of the Venturi has insignificant impact on the upstream flow as the outlet pipe is sufficiently long.

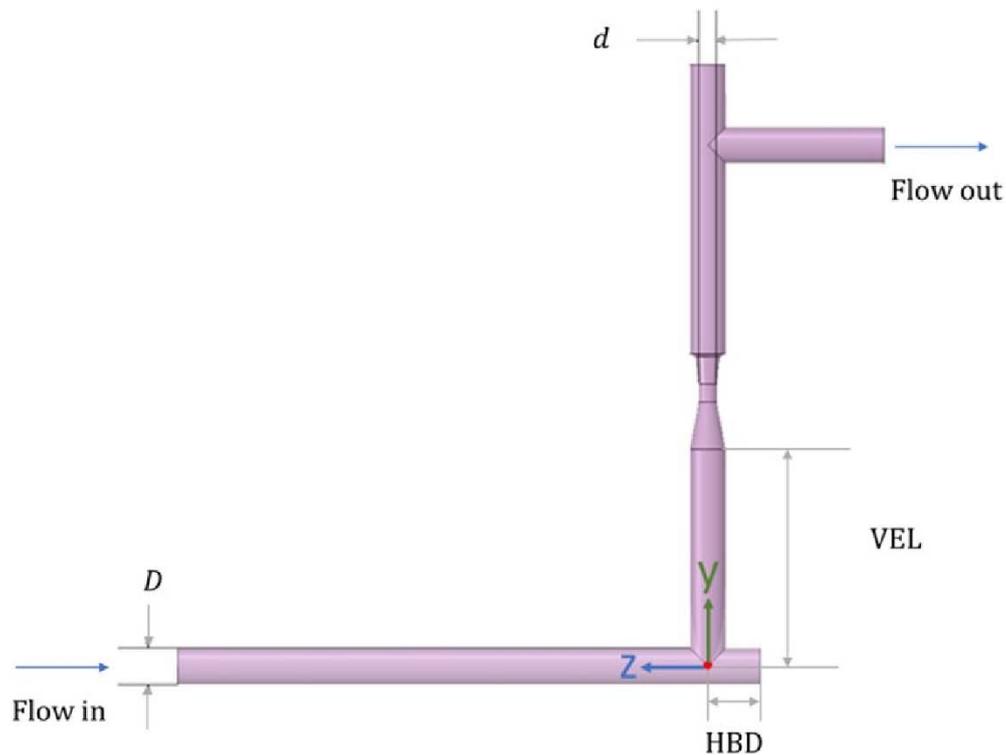

**Figure 1:** Geometry of three-dimensional computational model, with positive $x$-direction pointing out of plane.

Polyhedral mesh is used for its superior performance in accuracy and computational efficiency of other cell shapes such as tetrahedra (Milovan & Stephen, 2004). Simulation inputs include boundary conditions, material properties, and numerical schemes. Homogeneous phase velocity and fraction, turbulence intensity (~ 4% estimated), and pipe hydraulic diameter are specified at inlet conditions. Zero-gauge pressure is set at the outlet (assuming the flow has a negligible effect on the compressibility of the selected flow conditions). Material properties include the line-condition density and viscosity of the fluid. Calculating the interaction force also requires the surface tension of the liquid. The phase-coupled semi-implicit method for pressure linked equations (SIMPLE) algorithm and the first-order upwind scheme are used for spatial discretization. Mesh-independent results are used in the study. Our earlier work (Zhan *et al*., 2022) described the numerical model, mesh, and simulation setup in detail. Simulations are performed in the commercial CFD software Ansys Fluent 2021 R1.

## 2.2   Scope of study

Simulations are performed using ten test points for water-nitrogen flow (T1 to T5) and oil-nitrogen flow (T6 to T10) with $T \sim 30^{o}C$ and $P \sim 20bar$ from experimental multiphase flow-loops, covering a wide-range of flow inlet gas volume fractions (GVF$=\frac{\dot{Q}_g}{\dot{Q}_g+\dot{Q}_l}$, where $\dot{Q}_g$ and $\dot{Q}_l$ are the gas and liquid volumetric flow rates, respectively) from $\sim 26\%$ to $\sim 83\%$ and a series of flow inlets with homogeneous velocity from $1.68\,\mathrm{m/s}$ to $6.07\,\mathrm{m/s}$ (Figure 2(a)). Note that in the simulation setup, gas and liquid phases are uniformly distributed at the horizontal inlet, and the flow is allowed to develop along the horizontal pipe. Sensitivity studies have been performed to ensure that the horizontal entrance length used is sufficiently long for the flow to develop before entering the vertical pipe through the horizontal-to-vertical blind-tee. An indicative flow pattern mapping of the flow conditions at the horizontal inlet based on the Baker's horizontal flow regime map (Baker, 1953) is shown in Figure 2(b).

(a)

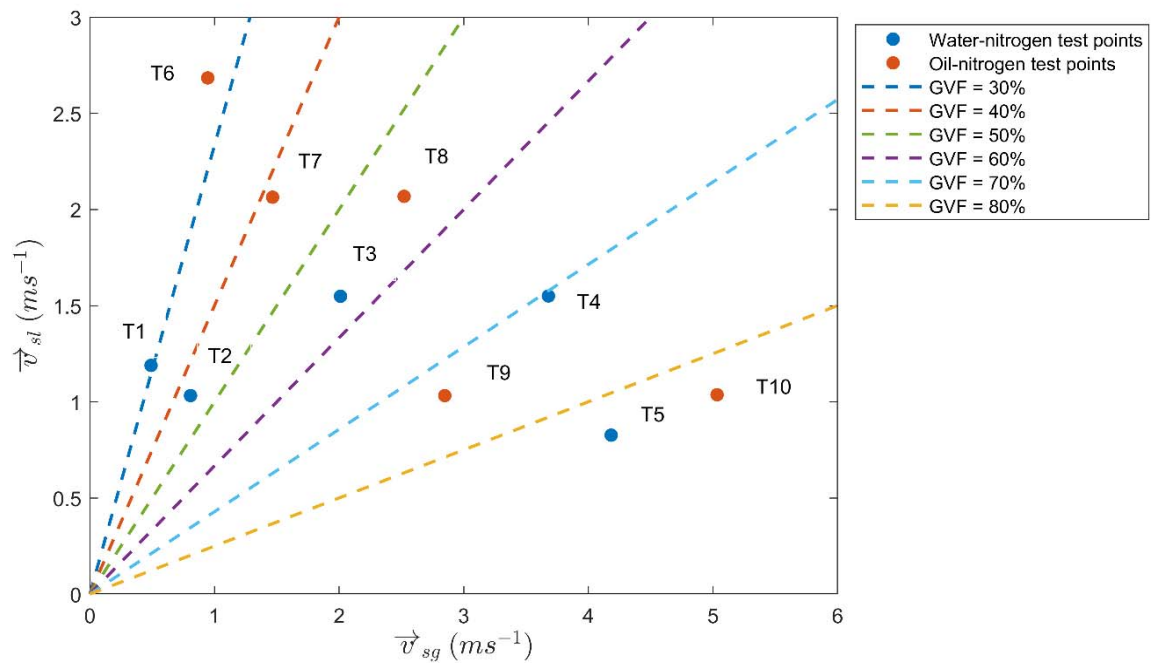

(b)

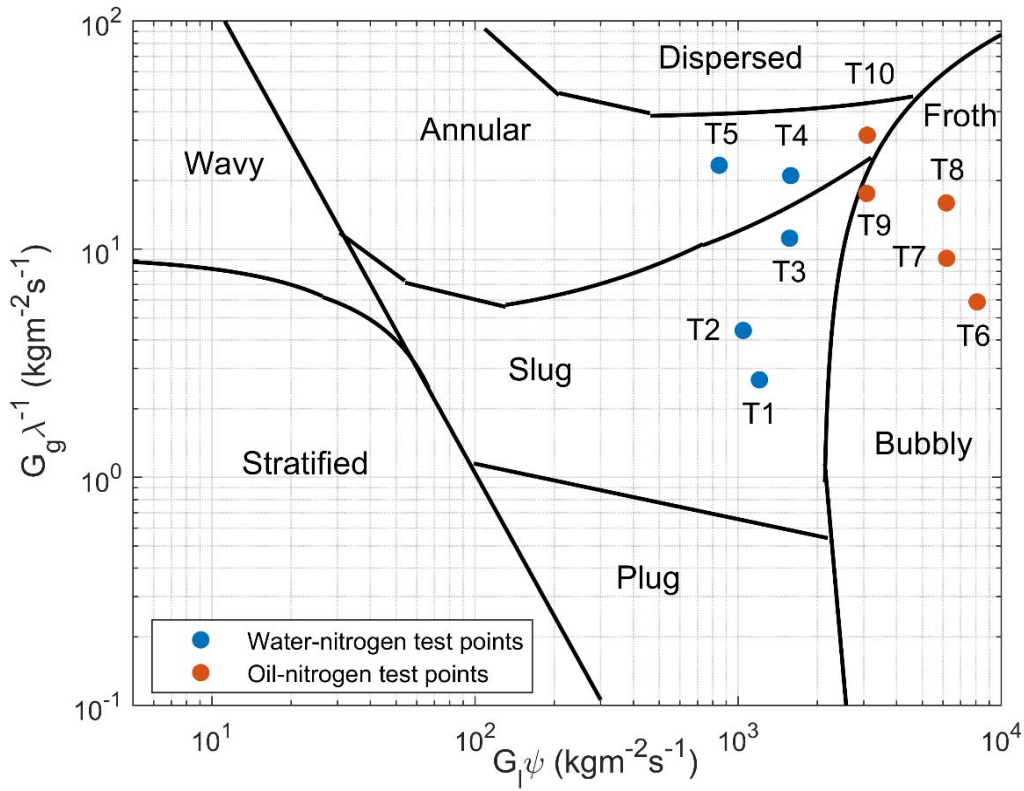

**Figure 2:** Flow conditions of water-nitrogen test points T1 to T5 and oil-nitrogen test points T6 to T10. $\vec{v}_{sg}$ is gas superficial velocity, $\vec{v}_{sl}$ is liquid superficial velocity, and $\vec{v}_{sg} + \vec{v}_{sl}$ is homogeneous velocity; (b) an indicative flow pattern mapping of flow conditions at a horizontal inlet based on Baker's horizontal flow regime map (Baker, 1953). $G_l$ is liquid mass flux, $G_g$ is gas mass flux. $\lambda = \sqrt{\dfrac{\rho_g}{\rho_{air}}\dfrac{\rho_l}{\rho_{water}}}$, $\psi = \dfrac{\sigma_{water}}{\sigma}\left(\dfrac{\mu_l}{\mu_{water}}\right)^{\frac{1}{3}}\left(\dfrac{\rho_{water}}{\rho_l}\right)^{\frac{2}{3}}$, where $\rho$, $\sigma$, and $\mu$ are fluid density, surface tension, and viscosity, respectively.

The simulation results are validated against the diametrically chord-averaged gas fraction $\alpha_{g,gamma}$ measured by the gamma-ray sensor in the $z$-direction (see Figure 1) and against the time-averaged Venturi $\Delta P$ measured between the mid-throat and the Venturi inlet (one pipe-diameter upstream of the convergent section). The agreement between the simulation and the experimental measurements is shown in Figure 3.

(a)

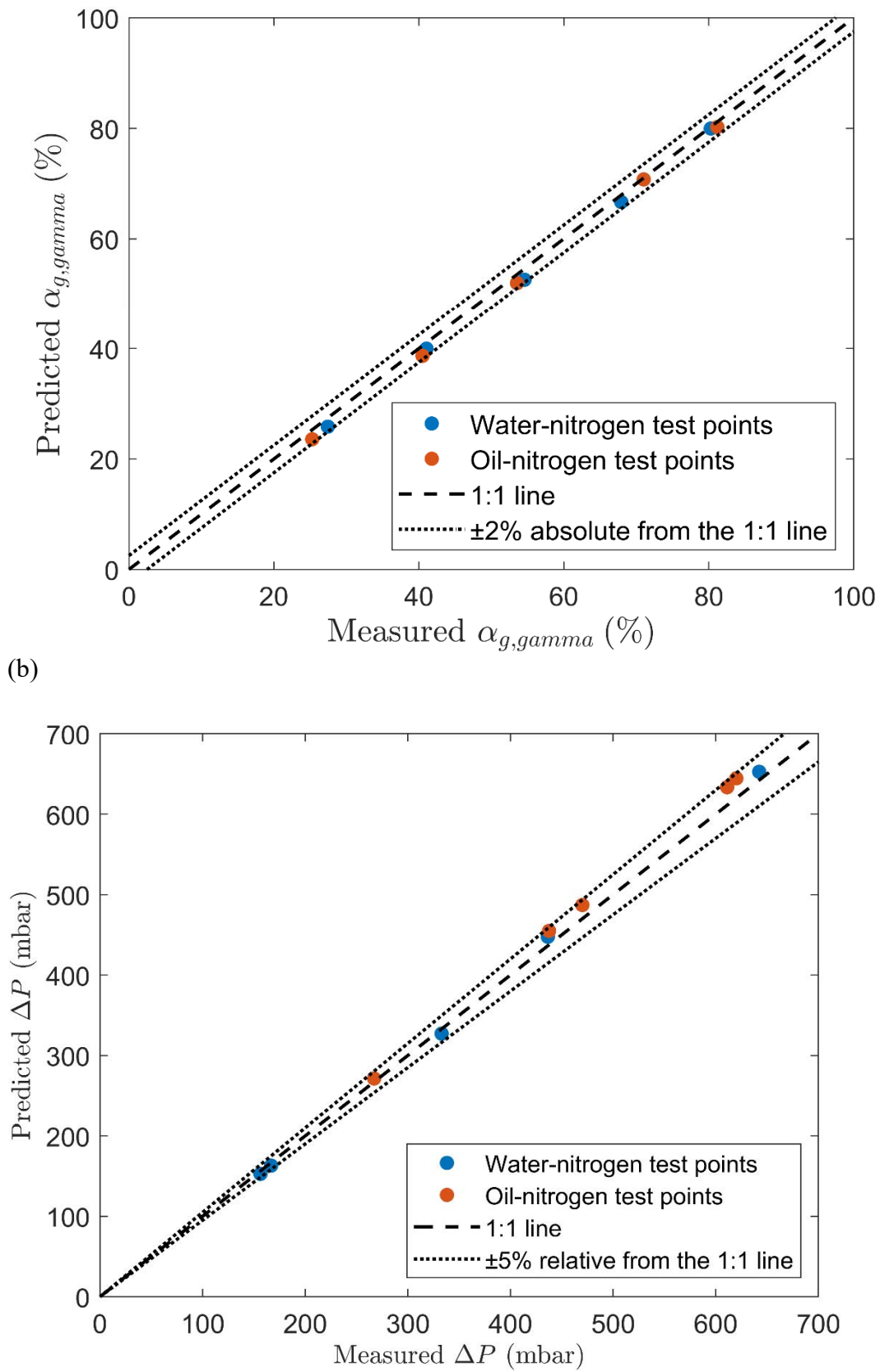

(b)

**Figure 3:** Computational fluid dynamics-predicted: (a) chord-averaged gas fraction $\alpha_{g,gamma}$ at throat against gamma-ray measurement and (b) Venturi $\Delta P$ against measurement.

For the absolute difference between predicted and measured chord-averaged gas fraction $\alpha_{g,gamma}$, all ten test points have a $\alpha_{g,gamma}$ difference $\leq 2\%$. In the terms of the Venturi $\Delta P$, there is a relative difference $\leq 5\%$ across all ten test points. Ten simulated test points are used for the study of HBD and VEL. The gas-liquid flow is numerically investigated in seven different flow domains, with their VEL and HBD shown in Table 1.

**Table 1:** Vertical entrance length (VEL) and horizontal blind-tee depth (HBD) for seven flow domains, where $D$ is inlet pipe-diameter (see Figure 1).

| Flow domain | VEL | HBD |
|:-----------:|:---:|:---:|
| A1 | $3D$ | |
| B1 | $6D$ | $1.5D$ |
| C1 | $11D$ | |
| D1 | $16D$ | |
| B2 | | $2D$ |
| B3 | $6D$ | $2.5D$ |
| B4 | | $3D$ |

## 3    Results and discussion

### 3.1    Study of vertical entrance length

For the study of VEL, the HBD used is $1.5D$ for all VEL simulations. The effect of different VELs on phase-fraction, differential-pressure and two-phase discharge coefficient is investigated.

#### 3.1.1    Phase-fraction

The variation of the cross-sectional gas fraction at the Venturi inlet $\alpha_{g,vi}$, the Venturi mid-throat $\alpha_{g,mt}$ and the gamma-beam-equivalent gas fraction $\alpha_{g,gamma}$ against GVF obtained from simulations of the flow domains A1, B1, C1, and D1 with increasing VEL is shown in Figures 4 and 5, respectively. Note that $\alpha_g$ and GVF can be related to the gas-liquid slip ratio $S = \frac{|\vec{v}_g|}{|\vec{v}_l|}$ (Shu *et al.*, 2016; 2017; 2018), as shown in Equation 1, where $\vec{v}_g$ and $\vec{v}_l$ are the actual velocities of gas and liquid, respectively.

$$\alpha_g = \frac{\text{GVF}}{S\left(1-\text{GVF}\right)+\text{GVF}}, \tag{1}$$

For vertical upward gas-liquid flow, mainly $S > 1$, $\alpha_g < \text{GVF}$; for non-slip gas-liquid flow, $S = 1$, $\alpha_g = \text{GVF}$; for vertical downward gas-liquid flow, mainly $S < 1$, $\alpha_g > \text{GVF}$.

(a)

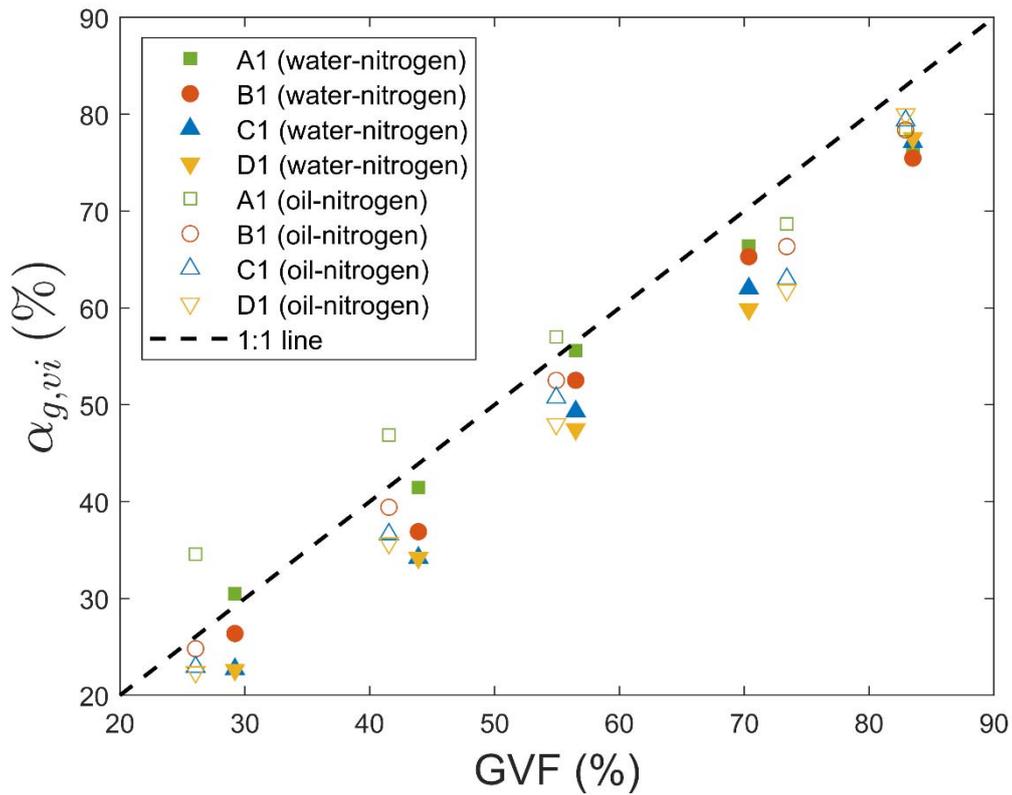

(b)

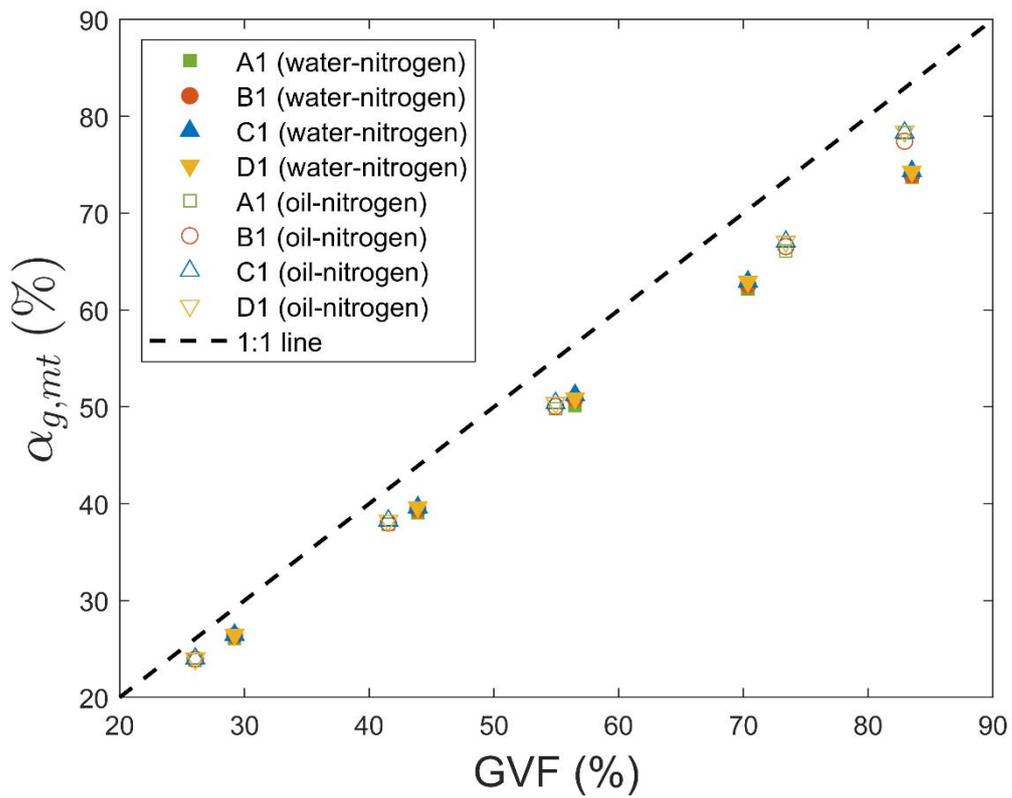

**Figure 4:** Variation of cross-sectional gas fraction with inlet gas volume fraction (GVF) at Venturi inlet $\alpha_{g,vi}$ (a) and Venturi mid-throat $\alpha_{g,mt}$ (b) for gas-liquid flow in flow domains A1, B1, C1, and D1.

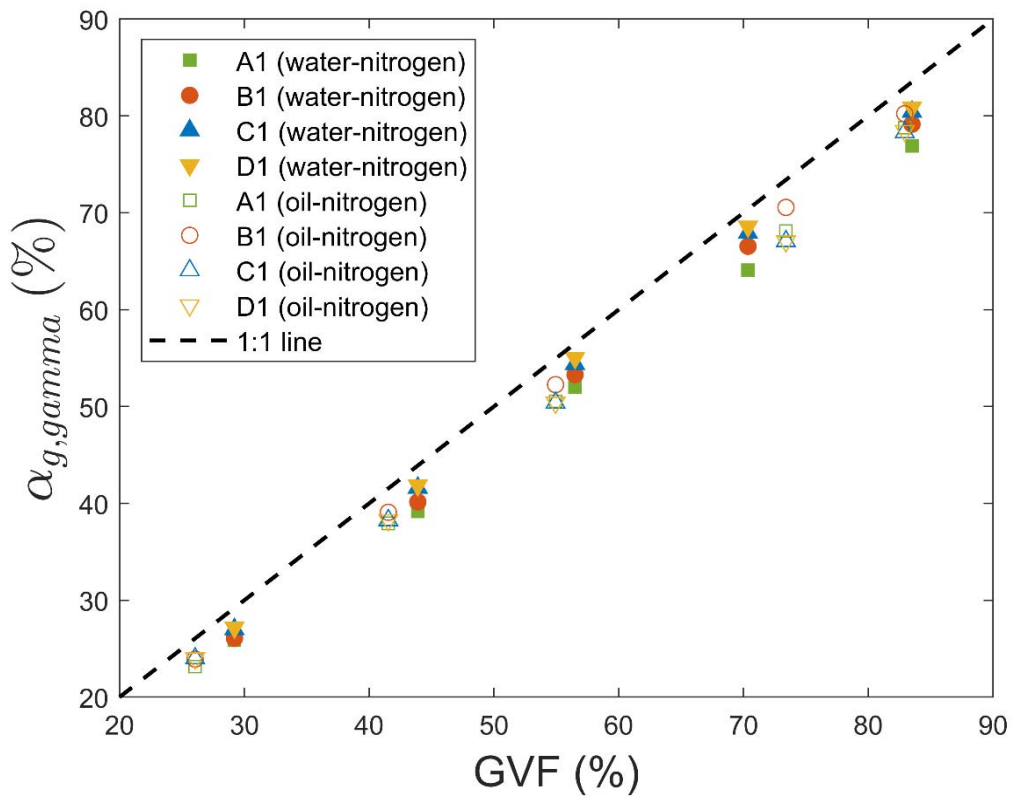

**Figure 5:** Variation of gamma-beam-equivalent gas fraction $\alpha_{g,gamma}$ at Venturi throat along $z$-direction with inlet gas volume fraction (GVF) for gas-liquid flow in flow domains A1, B1, C1, and D1.

It is observed from Figure 4 that at the Venturi inlet, $\alpha_{g,vi}$ decreases with increasing VEL in a range of $20\% \leq GVF \leq 80\%$. The difference in $\alpha_{g,vi}$ between the gas-liquid flow with the shortest VEL in A1 ($3D$) and the longest VEL in D1 ($6D$) ranges from 6.54% to 12.21%. According to mass flow conservation, smaller gas fractions indicate larger gas velocities and larger gas-liquid slip $S$, for a given homogeneous inlet velocity. Hence, the gas-liquid flow in C1 and D1 has greater gas-liquid slip $S$ than in A1 and B1. Note that most of the test points have $\alpha_{g,vi}$ values below GVF, suggesting that gas has a higher velocity than the liquid or the gas-liquid slip $S$ greater than 1; however, below a certain GVF ($60\%$ for oil-nitrogen flow and $30\%$ for water-nitrogen flow), $\alpha_{g,vi}$ in A1 is greater than GVF, suggesting that gas has a lower velocity than the liquid or the gas-liquid slip $S$ smaller than 1 at the Venturi inlet in A1. This may be because there is a greater VEL in C1 and D1 than in A1 and B1 for gas-liquid slip to develop after the flow from the horizontal pipe and the horizontal blind-tee into the vertical pipe. For $GVF \geq 80\%$, the difference in $\alpha_{g,vi}$ of gas-liquid flow for different VELs is small, ranging from 0.99% to 1.59% between A1 and D1, suggesting that VEL has a small impact on the development of gas-liquid slip $S$ at high GVF. At the Venturi mid-throat, the difference in $\alpha_{g,mt}$ of gas-liquid flow for different VELs is insignificant, with a maximal $\alpha_{g,mt}$ difference between A1 and D1 being 0.96%. This suggests that VEL has a small impact on the development of gas-liquid slip $S$ in the Venturi throat section.

To gain a better understanding of the blind-tee homogenization effect, the evolution of the slip ratio $S$ is evaluated in six cross-sections, at $5D$ upstream of the vertical pipe centerline (in the horizontal pipe), $2D$, $5D$, $11D$ and $15D$ downstream of the horizontal pipe centerline (at the vertical Venturi inlet for A1, B1, C1, D1, respectively) and the Venturi mid-throat for D1. Note that at $5D$ upstream the vertical pipe centerline, the flow is developed with a gas fraction having a difference of 0.01% absolute with that at $10D$ upstream. Figure 6 shows the evolution of slip ratio $S$ for T1, T8, and T10 that may correspond to slug, bubbly, and annular flow in horizontal pipe, respectively (Figure 2(b)).

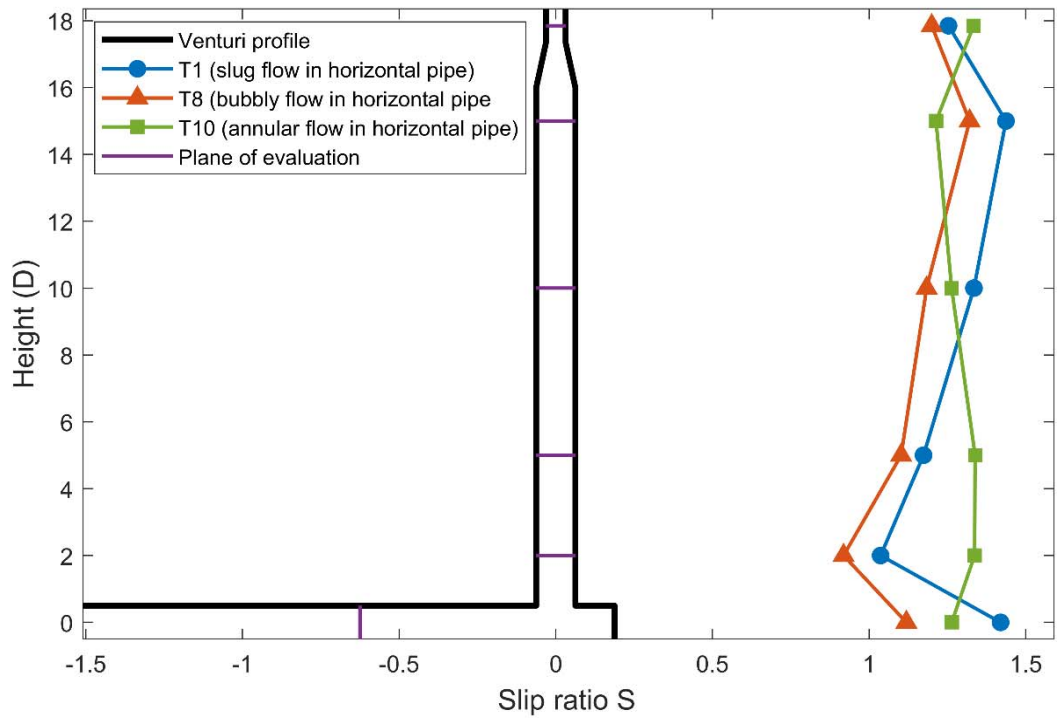

**Figure 6:** Evolution of slip ratio $S$ from $5D$ upstream vertical pipe centerline (in horizontal pipe) to $2D$, $5D$, $11D$ and $15D$ downstream horizontal pipe centerline and Venturi mid-throat for D1 for T1 (slug flow in horizontal pipe), T8 (bubbly flow in horizontal pipe), and T10 (annular flow in horizontal pipe).

Figure 6 shows that for slug flow (T1) and bubbly flow (T8), the homogenization effect of the

blind-tee is observed within $2D$ downstream of the horizontal pipe centerline, where the slip ratio

$S$ decreases from 1.4 and 1.1 at $5D$ upstream of the horizontal pipe to 1.1 and 0.9 at $2D$

downstream of the horizontal pipe centerline, respectively; however, the slip ratio $S$ increases

along the vertical pipe, as the flow develops. At $15D$ downstream of the horizontal pipe

centerline, the slip ratio $S$ increases to 1.4 and 1.3 before decreasing to 1.3 and 1.2 for T1 and

T8, respectively. The reduction at the Venturi mid-throat may be due to flow acceleration improving the mixing of the phases at the Venturi throat. Unlike dispersed and intermittent flow, the slip ratio $S$ of annular flow (T10) does not vary as much as that of slug and bubbly flow, with the slip ratio varying between 1.2 and 1.3. Note that slip increases at the Venturi throat, probably because the liquid layer closer to the wall encounters greater friction with the wall and accelerates to a smaller extent while the gas core accelerates to a greater extent. Hence, the homogenization of the horizontal blind-tee is most effective for the bubbly flow and the slug flow within $2D$ downstream of the horizontal pipe centerline, and has an insignificant effect on the slip ratio $S$ in the annular flow; however, the homogenization effect of the blind-tee does not seem to affect the Venturi throat where phase-fraction measurement is taken, as observed by similar $\alpha_{g,mt}$ with different VEL.

However, despite having the similar $\alpha_{g,mt}$, the gamma-beam-equivalent gas fraction $\alpha_{g,gamma}$ in the Venturi throat along the $z$-direction increases with VEL in a converging trend from A1 to D1, as shown in Figure 5. This is likely due to the phase distribution in the Venturi throat. To gain a deeper understanding of the phase distribution at different VELs, two examples of the phase distribution at the Venturi inlet and the Venturi mid-throat for gas-liquid flow in A1 to D1 are shown in Figure 7(a) and 7(b), for low-GVF water-nitrogen flow (T1) and high-GVF oil-nitrogen flow (T10), respectively.

(a)

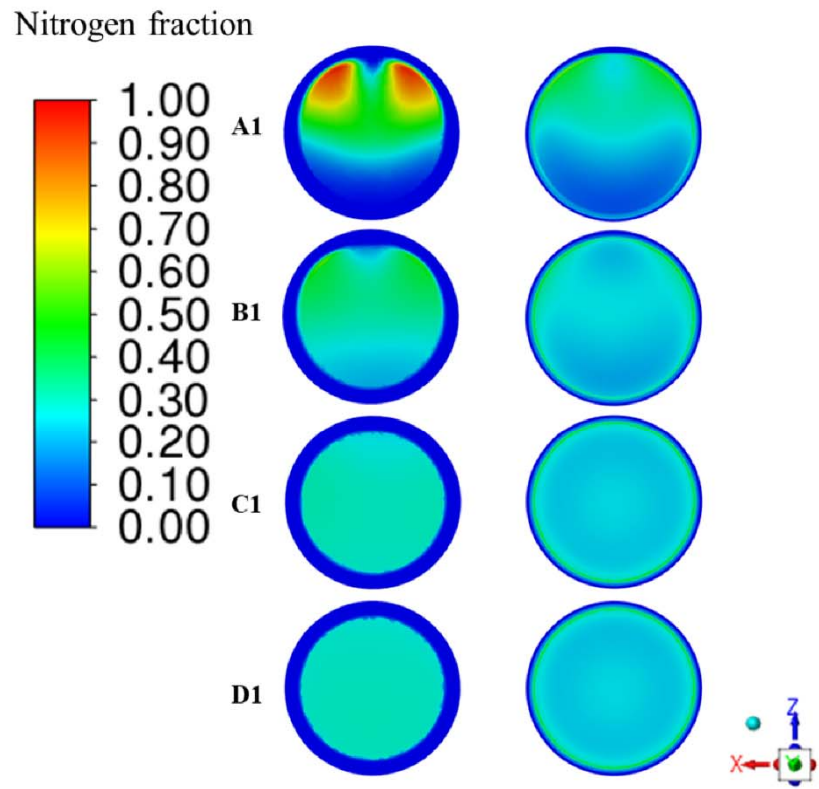

(b)

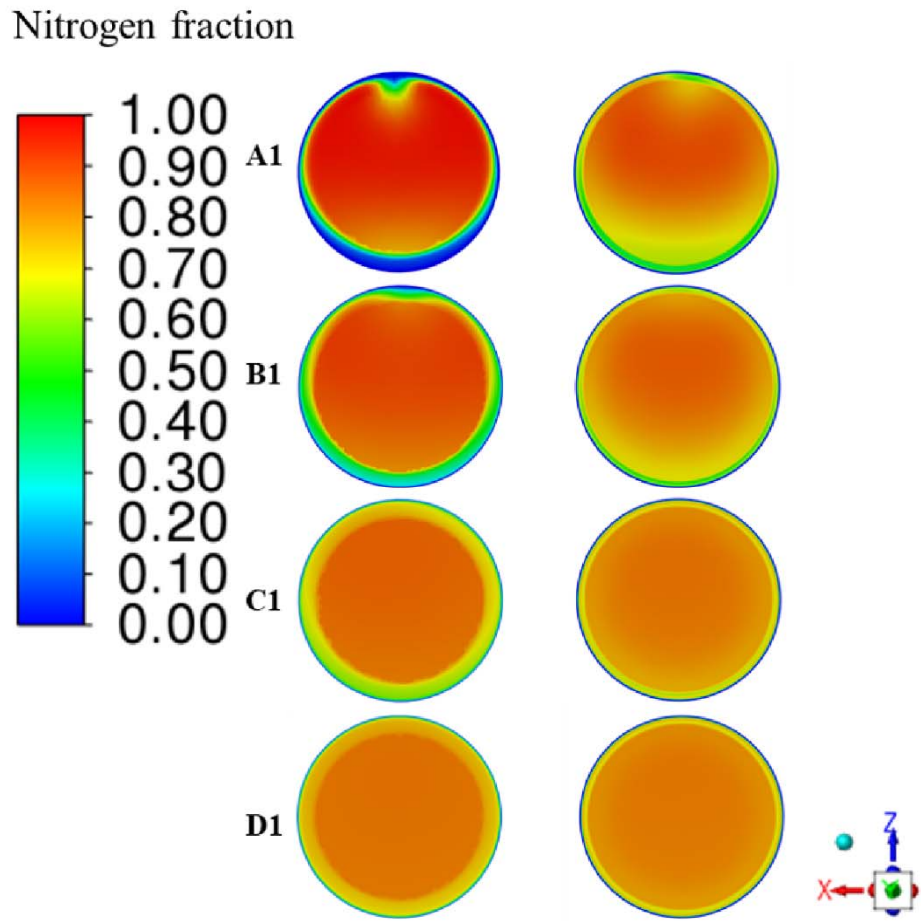

**Figure 7:** Cross-sectional gas fraction distribution of A1, B1, C1, and D1 at Venturi inlet (left column) and Venturi mid-throat (right column) for (a) water-nitrogen flow with gas volume fraction (GVF) 29%, homogeneous inlet velocity 1.68 m/s (T1) and (b) oil-nitrogen flow with GVF 83%, homogeneous inlet velocity 6.07 m/s (T10).

It is observed from Figure 7 that at the Venturi inlet, the phase distribution is more liquid-rich in the $-z$ half than the $+z$ half, despite there are some liquid-rich spots in the $+z$-direction along the $z$ diameter; however, the phase distribution is more axisymmetric at the Venturi mid-throat than at the Venturi inlet, due to the higher flow velocity and greater mixing at the Venturi throat. The phase distribution becomes more axisymmetric with an increase in VEL. The phase distribution is virtually axisymmetric at C1 and D1 at both the Venturi inlet and the Venturi mid-throat. This observation is consistent with the finding that the difference in $\alpha_{g,gamma}$ between C1 and D1 is insignificant. It can be interpreted that the VEL has an impact on the axisymmetry of the phase distribution at the Venturi inlet, which may affect the axisymmetry of the phase distribution and the directional chordal phase-fraction measured by the gamma-ray sensor located downstream of the Venturi throat. To quantify the effect of the axisymmetry of the phase distribution on the directional chordal phase-fraction measurement, the transverse gas fraction distribution and the mean transverse gas fraction along the $x$-direction and the $z$-direction at the Venturi mid-throat are shown in Figure 8 for water-nitrogen flow T1 and oil-nitrogen flow T10.

(a)

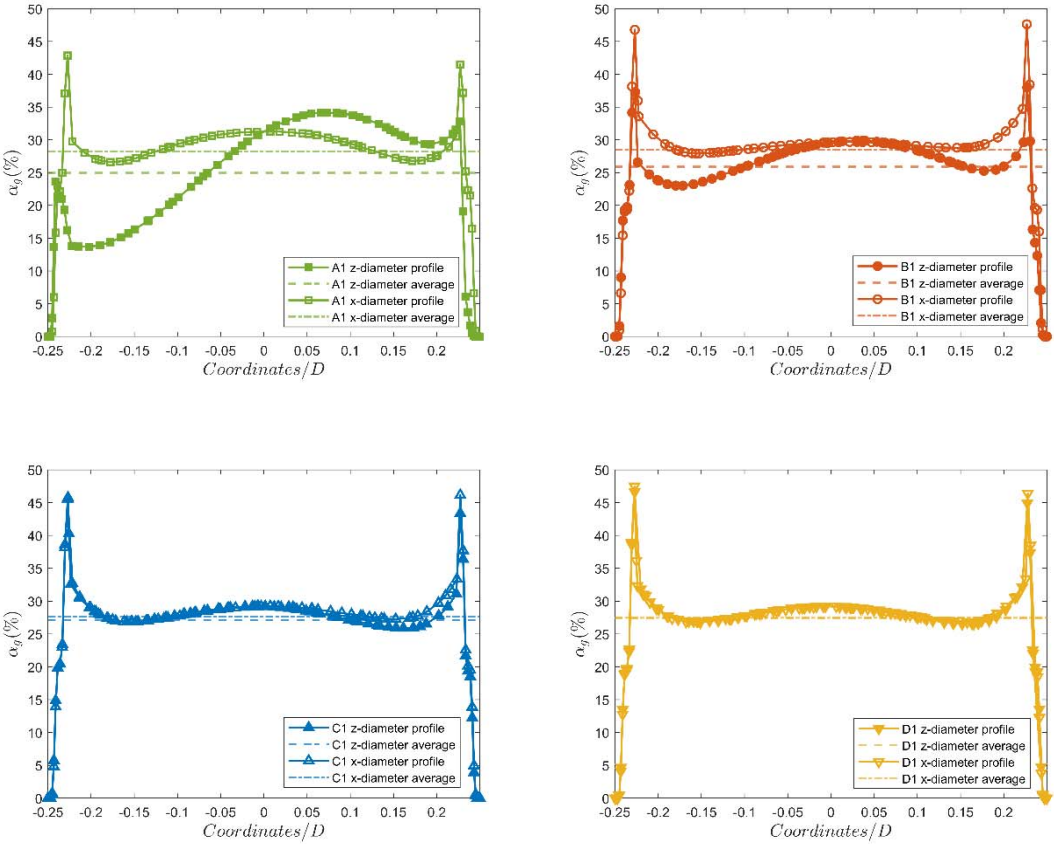

(b)

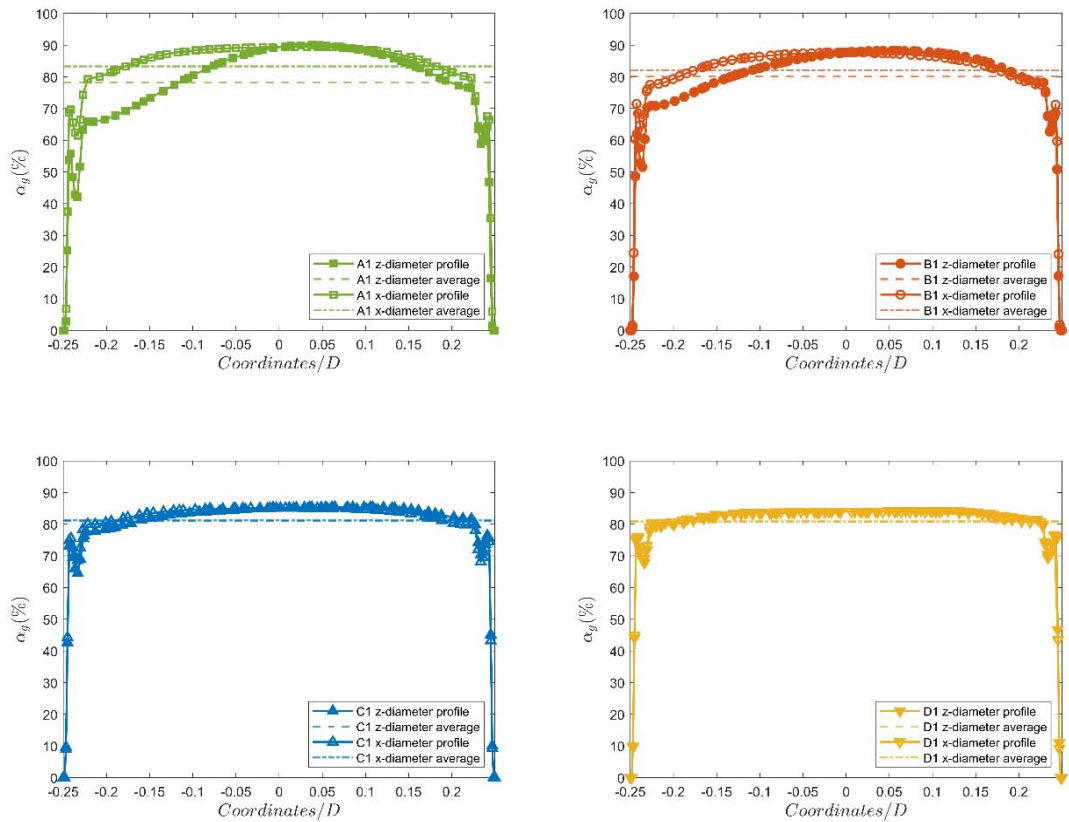

**Figure 8:** Transverse gas fraction distribution of A1, B1, C1, and D1 at Venturi mid-throat for (a) water-nitrogen flow with gas volume fraction (GVF) 29% , homogeneous inlet velocity 1.68 m/s (T1), and (b) oil-nitrogen flow with GVF 83% , homogeneous inlet velocity 6.07 m/s (T10).

Figure 8 indicates that while the transverse gas distribution profiles along the $x$ axis are symmetric about $x = 0$ , the phase distribution along the $z$ -direction is more liquid-rich at the $-z$ -direction for A1 and B1. The difference in $x$ and $z$ gas distribution profiles and the gas fraction mean decreases with an increase in VEL and the axisymmetry of the phase distribution. In addition, the difference decreases faster in T10 than T1. For T1, the difference in $x$ and $z$ gas fraction mean decreases from 3.28% in A1, to 2.64% in B1, 0.52% in C1, and 0.05% in D1. For T10, the difference in $x$ and $z$ phase-fraction mean decreases from 5.1% in A1, to 1.87% in B1, 0.18% in C1, and 0.04% in D1. This may be because the flow in T10 has a greater homogeneous inlet velocity than T1, which results in greater mixing and allows axisymmetry to be reached at lower VEL.

### 3.1.2  *Venturi differential-pressure*

Figure 9 shows the simulated differential-pressure between the Venturi mid-throat and the Venturi inlet for all test points in flow domains A1, B1, C1, and D1. Note that the distance between the Venturi inlet and the Venturi mid-throat is the same for all flow domains.

(a)

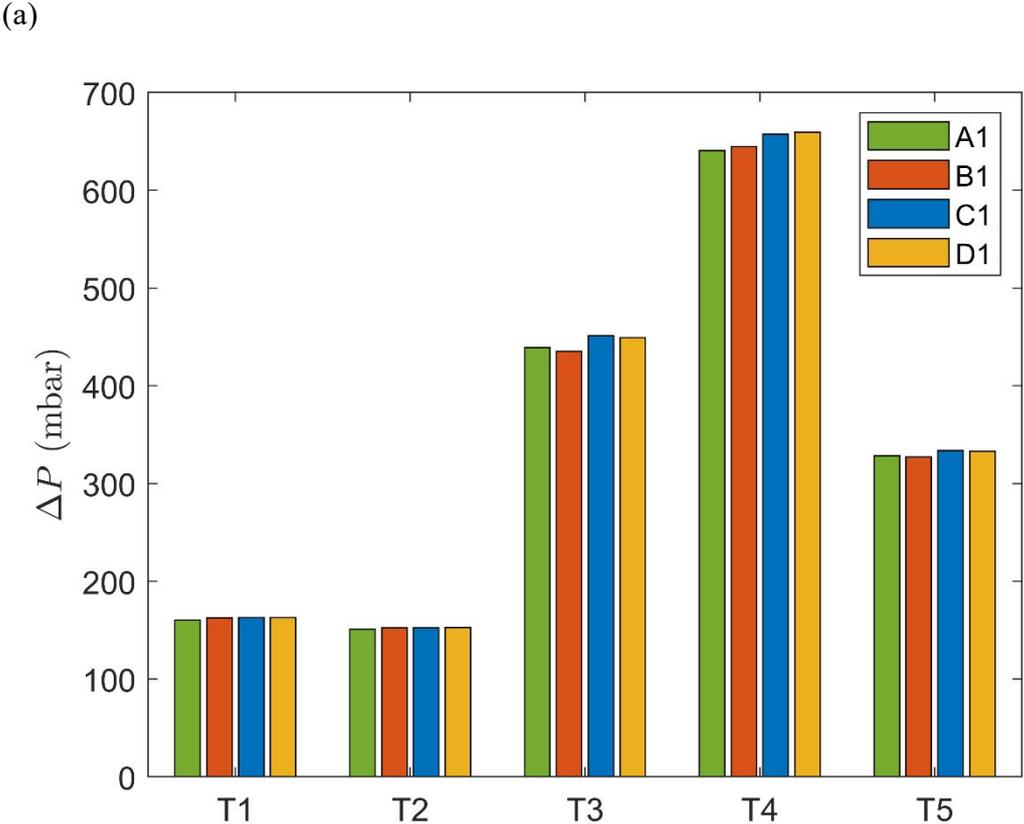

(b)

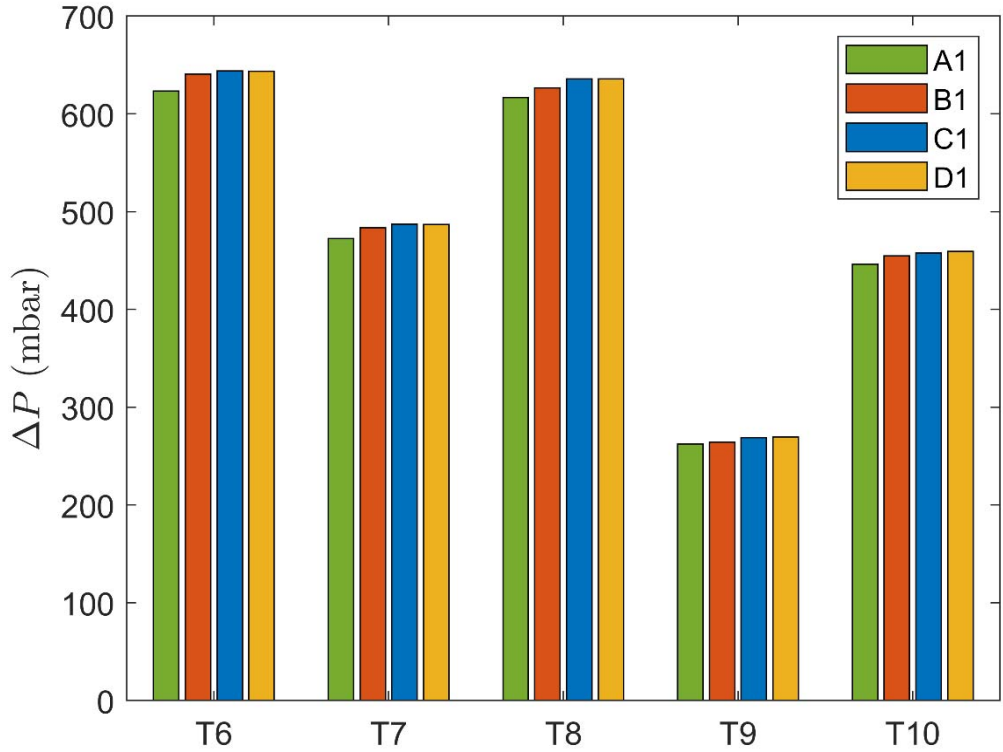

**Figure 9:** Simulated Venturi $\Delta P$ for flow domains A1, B1, C1, and D1, for (a) water-nitrogen test points (T1–T5); and (b) oil-nitrogen test points (T6–T10).

As can be seen in Figure 9, the flow in C1 and D1 has higher $\Delta P$ than in A1 and B1. As can be seen from Figure 3, the difference in the cross-sectional gas fraction $\alpha_{g,mt}$ at the Venturi mid-throat in the gas-liquid flow with increasing VEL (A1, B1, C1, and D1) is negligible, while the phase-fraction at the Venturi inlet $\alpha_{g,vi}$ decreases with increasing VEL. Due to the conservation of mass flow, the liquid velocities at the Venturi inlet $\vec{v}_{l,vi}$ in A1 and B1 are greater than those in C1 and D1, while the liquid velocities at the Venturi throat $\vec{v}_{l,mt}$ are similar for the gas-liquid flow with different VELs. According to the Bernoulli equation given in Equation 2, the largest amount of work done by the pressure is to increase the kinetic energy of the liquid, given that the liquid density is about two orders of magnitude higher than the gas density. Hence, a greater differential-pressure is needed to accelerate the liquid in C1 and D1 from a lower liquid velocity $\vec{v}_{l,vi}$ at the Venturi inlet than in A1 and B1 to similar liquid velocity $\vec{v}_{l,mt}$ at the Venturi throat.

$$\Delta P = \alpha_{g,mt}\rho_{g,mt}gh_{mt} + \frac{1}{2}\alpha_{g,mt}\rho_{g,mt}\left|\vec{v}_{g,mt}\right|^2 + \alpha_{l,mt}\rho_{l,mt}gh_{mt} + \frac{1}{2}\alpha_{l,mt}\rho_{l,mt}\left|\vec{v}_{l,mt}\right|^2$$

$$-\alpha_{g,vi}\rho_{g,vi}gh_{vi}-\frac{1}{2}\alpha_{g,vi}\rho_{g,vi}\left|\vec{v}_{g,vi}\right|^2-\alpha_{l,vi}\rho_{l,vi}gh_{vi}-\frac{1}{2}\alpha_{l,vi}\rho_{l,vi}\left|\vec{v}_{l,vi}\right|^2+\text{Frictional Loss.}$$

$$(2)$$

To quantify the effect of VEL on the differential-pressure, the relative difference in differential-pressure with respect to D1 $\dfrac{\Delta P(\text{D1})-\Delta P(\text{A1}/\text{B1}/\text{C1})}{\Delta P}$ is shown in Figure 10.

(a)

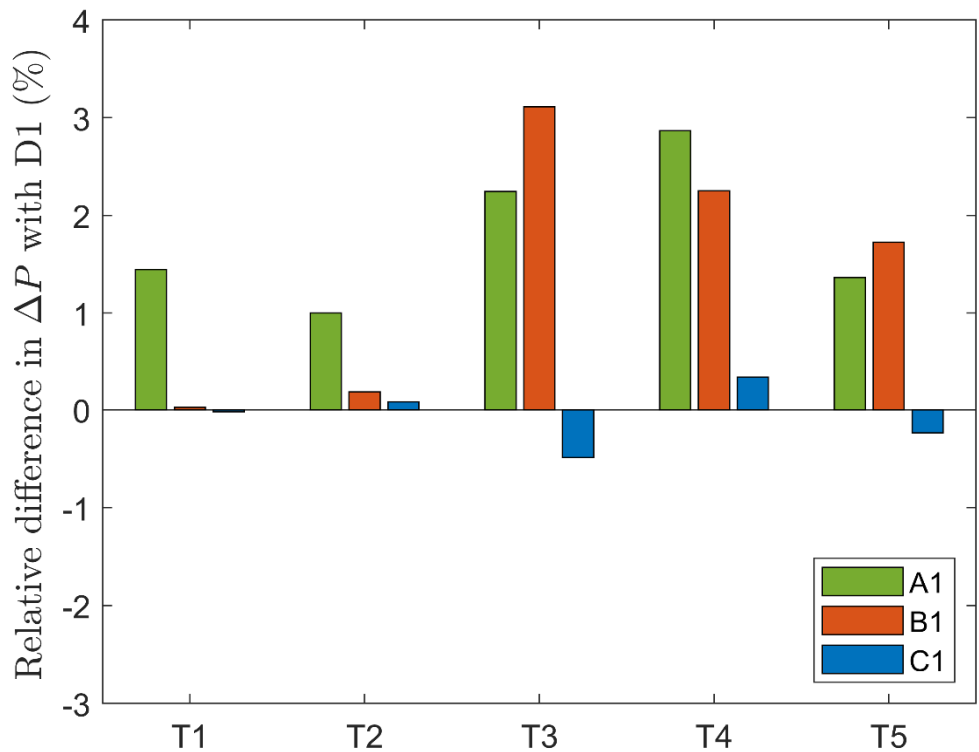

(b)

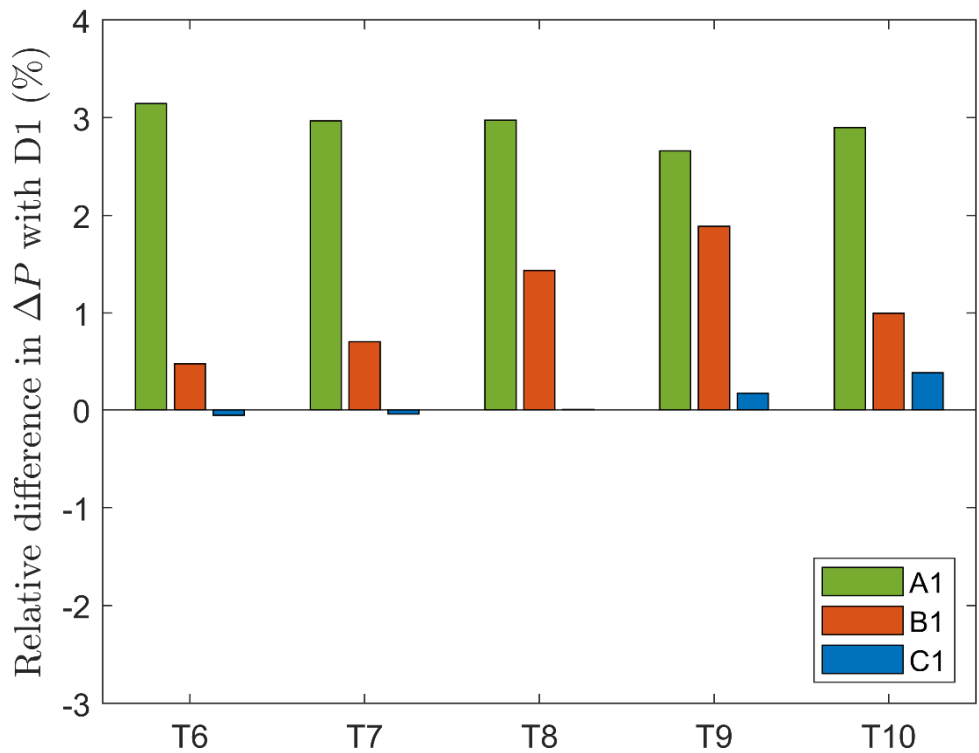

**Figure 10:** Relative difference in $\Delta P$ with respect to flow domain D1 for (a) water-nitrogen test points (T1 to T5); and (b) oil-nitrogen test points (T6 to T10).

As can be seen from Figure 10, in general, A1 has the largest relative difference in $\Delta P$ with respect to D1, and the relative difference ranges from 1% to 2.87% in water-nitrogen flow (T1 to T5), and from 2.66% to 3.14% in oil-nitrogen flow (T6 to T10). B1 has a smaller relative difference in $\Delta P$ with respect to the D1, and the relative difference ranges from 0.03% to 3.11% in water-nitrogen flow and from 0.48% to 1.89% in oil-nitrogen flow. C1 has the smallest relative difference in $\Delta P$ with respect to D1, and the relative difference is less than $\pm 0.5\%$ for all water-nitrogen and oil-nitrogen test points. Overall, the effect of VEL on the Venturi $\Delta P$ is insignificant, with a maximum relative difference of ~ 3% among different VELs.

### 3.1.3 Two-phase discharge coefficient

In practice, the measured differential-pressure is often greater than that estimated by the Bernoulli's principle with the assumption of no energy loss, because more energy is required to account for, for example, friction-induced energy losses in a Venturi. Hence, the mass flow rate $\dot{q}_{m,raw}$ given in Equation 3 is likely to be overestimated. Therefore, to account for energy losses in the Venturi, the discharge coefficient is often used to correct for the raw mass flow rate $\dot{q}_{m,raw}$. For single-phase flows, the discharge coefficient $C_d$ is found to be related to the Reynolds number $Re$ of the flow. The $C_{d(\dot{m})} - Re$ curve derived experimentally from a single-phase liquid in a Venturi is shown in Figure 11. The $C_{d(\dot{m})}$ is defined as the coefficient needed to correct the overpredicted mass flow rate under the assumption of no energy loss using Equation 4.

$$\dot{q}_{m,raw} = \frac{1}{\sqrt{1-\beta}} \frac{\pi}{4} d^2 \sqrt{2\rho_f \left( \Delta P - 2\rho_f g \Delta h \right)}, \tag{3}$$

where the hydrostatic pressure height $\Delta h = h_{mt} - h_{vi}$, $\rho_f$ is the fluid density and $\beta = \dfrac{d}{D}$ .

$$\dot{q}_{tm,c} = C_{d(\dot{m})} \dot{q}_{m,raw}, \tag{4}$$

where $\dot{q}_{tm,c}$ is the true mass flow rate.

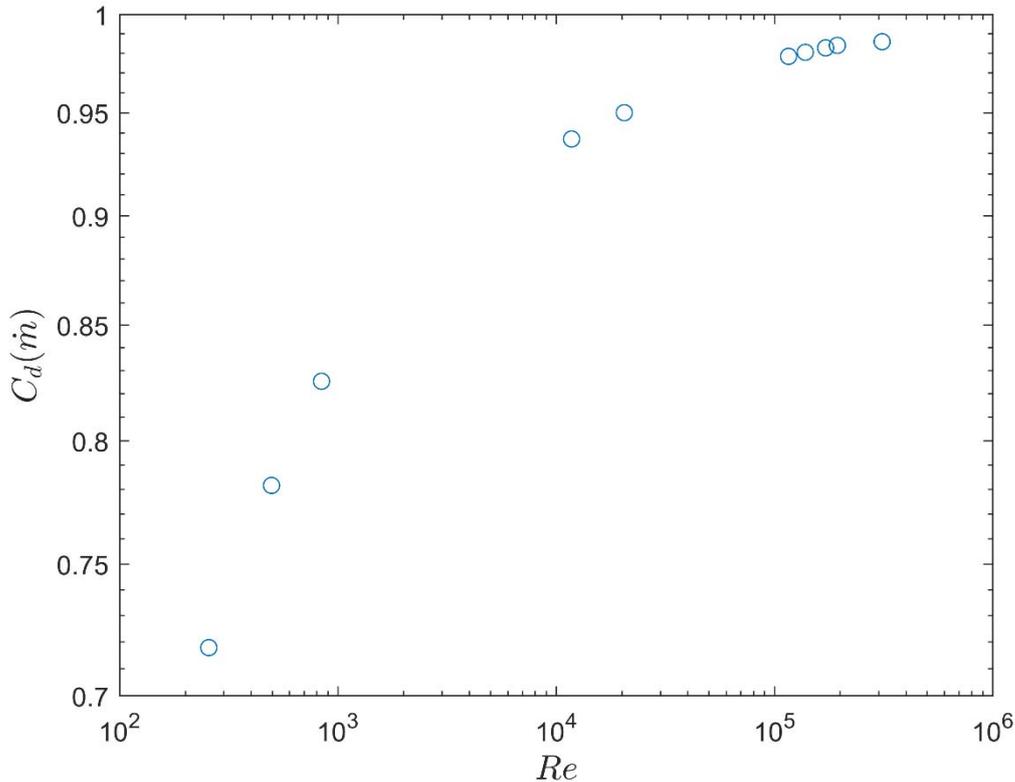

**Figure 11:** Single-phase discharge $C_{d(\dot{m})}$ coefficient of Venturi meter (Figure 1) obtained experimentally.

From Figure 11, it is observed that $C_d$ increases from approximately 0.72 to 0.99 with an increasing $Re$ number from $\sim 10^2$ to $\sim 10^5$.

The two-phase Venturi discharge coefficient $C_{d,tp}$ is used to account for losses in two-phase flow. $C_{d,tp}$ is evaluated for all test points of A1, B1, C1, and D1. In this study, $C_{d,tp}$ is defined as the ratio of the sum of change of the kinetic energy $KE$ and the gravitational potential energy $UE$ to the work done in the $y$-direction (see Figure 1) $W_y$ across the Venturi inlet and the Venturi mid-throat, $C_{d,tp} = \dfrac{\Delta KE + \Delta UE}{\Delta W_y}$, where $\Delta KE = KE_{mt} - KE_{vi}$, $\Delta UE = UE_{mt} - UE_{vi}$, and $\Delta W_y = \left| W_{y,mt} - W_{y,vi} \right|$. The kinetic energy $KE_p$, the gravitational potential energy $UE_p$, and the work done $W_{y,p}$ are respectively evaluated using Equation 3 over the plane of the Venturi inlet and the Venturi mid-throat. Note that only $v_y$ is considered in the evaluation as $v_y$ is at least two orders higher in magnitude than $v_x$ and $v_z$. $W_{y,p} = \delta t \sum_{i=1}^{n} \left( \alpha_{g,i} v_{gy,i} + \alpha_{l,i} v_{ly,i} \right) P_{y,i} A_i$,

$KE_p = \dfrac{1}{2} \delta t \sum_{i=1}^{n} \left( \rho_{g,i} \alpha_{g,i} v_{gy,i}^3 + \rho_{l,i} \alpha_{l,i} v_{ly,i}^3 \right) A_i$, $UE_p = gh \delta t \sum_{i=1}^{n} \left( \rho_{g,i} \alpha_{g,i} v_{gy,i} + \rho_{l,i} \alpha_{l,i} v_{ly,i} \right) A_i$, where $\delta t$ is the time interval, and the subscript $i$ stands for the $i$th grid in the mesh of the plane, and $A_i$ is the area of the $i$th grid in the mesh of the plane. Figure 12 shows that the two-phase discharge coefficient $C_{d,tp}$ increases with increasing VEL (A1, B1, C1, and D1) in a converging trend. A1 has the smallest VEL and the most asymmetric phase distribution, and is observed to have the lowest $C_{d,tp}$ or greatest loss at all test points, while C1 and D1, which have the most axisymmetric phase distribution, are observed to have the highest $C_{d,tp}$ and smallest loss for most test points (except for T3 where B1 has a higher $C_{d,tp}$ than D1). B1 significantly has a lower $C_{d,tp}$ than C1 and D1; however, the difference is insignificant for most test points, with the maximal absolute difference of 0.6% for D1 in T10, compared to the difference of 1.9% between A1 and D1 in T10. The trend is consistent with the gamma-beam-equivalent gas fraction $\alpha_{g,gamma}$ (see Figure 5) and the degree of axisymmetry of the phase distribution at the Venturi inlet and the Venturi mid-throat (see Figure 7). Hence, it can be deduced that a flow with a more axisymmetric phase distribution may suffer less loss than a flow with a less axisymmetric phase distribution. This may be because the gas-liquid flow with a less axisymmetric phase distribution has a greater near-wall liquid content, particularly in the $-z$-direction (see Figure 7), resulting in a greater loss due to the friction between the pipe wall and the near-wall liquid layer.

(a)

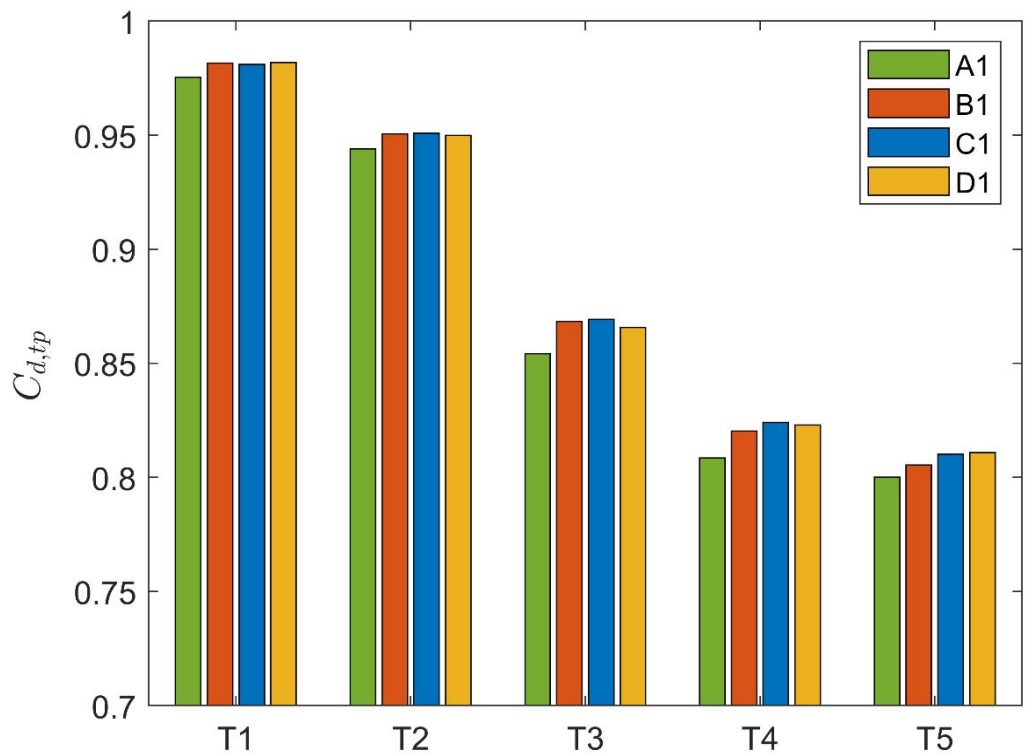

(b)

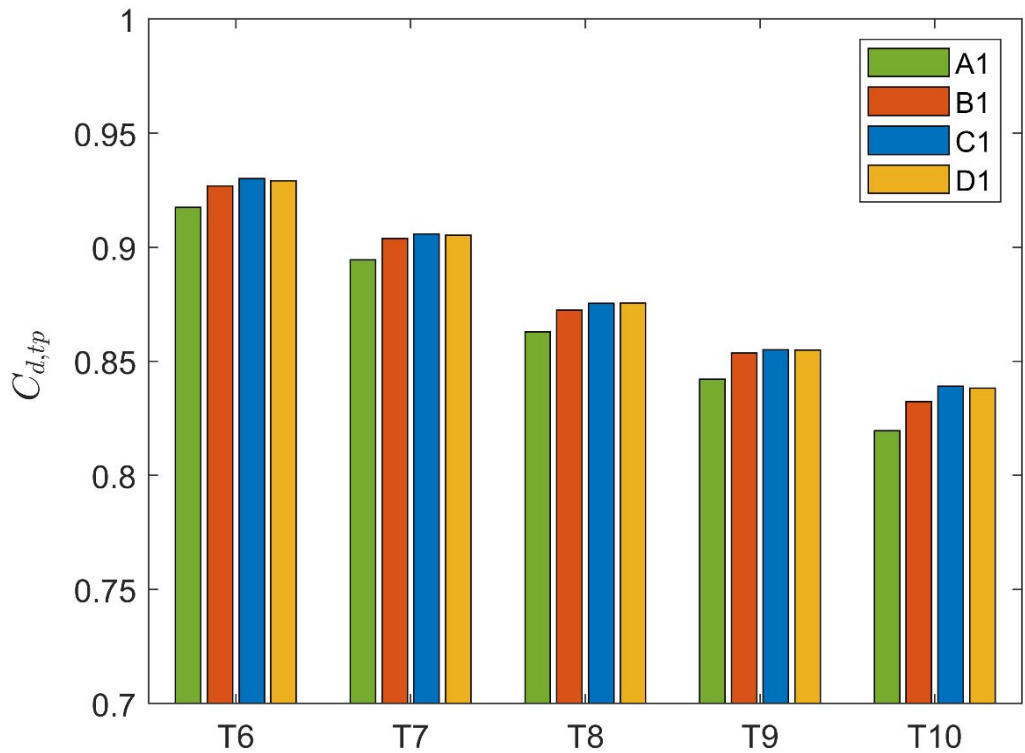

**Figure 12:** Two-phase discharge coefficient $C_{d,tp}$ for flow domains A1, B1, C1, and D1, for (a) water-nitrogen test points (T1 to T5); and (b) oil-nitrogen test points (T6 to T10).

From the study, since the agreement between the simulated $\alpha_{g,gamma}$ and the measured $\alpha_{g,gamma}$ is within 2%, flow domain B1 (which has a less than 2% difference in $\alpha_{g,gamma}$ with respect to flow domain D1) is considered suitable for MPFM with a minimal VEL being $6D$. The difference in $\Delta P$ for all flow domain geometries are less than 5%. Considering the relative difference between simulated and measured $\Delta P$ is less than 5%. The differences in $\Delta P$ of flow domains with different VELs can be considered insignificant. The two-phase discharge coefficients of the gas-liquid flow in B1, C1, and D1 are also greater than those in A1. In order to maintain the multiphase flow measurement accuracy within 2.5 to 5%, and minimize the energy loss, manufacturing cost, and carbon footprint, in this study, VEL of $6D$ is considered suitable for a vertically mounted Venturi-based MPFM.

### 3.2    Study of horizontal blind-tee depth

A fixed VEL of $6D$ is used to study the effect of HBD on the measurements of phase-fraction, differential-pressure, and local liquid-property.

#### 3.2.1    Phase-fraction

The cross-sectional gas fraction of the Venturi mid-throat $\alpha_{g,mt}$ and the gamma-beam-equivalent gas fraction $\alpha_{g,gamma}$ obtained from simulations with increasing HBD (B1, B2, B3, and B4) is shown in Figure 13.

(a)

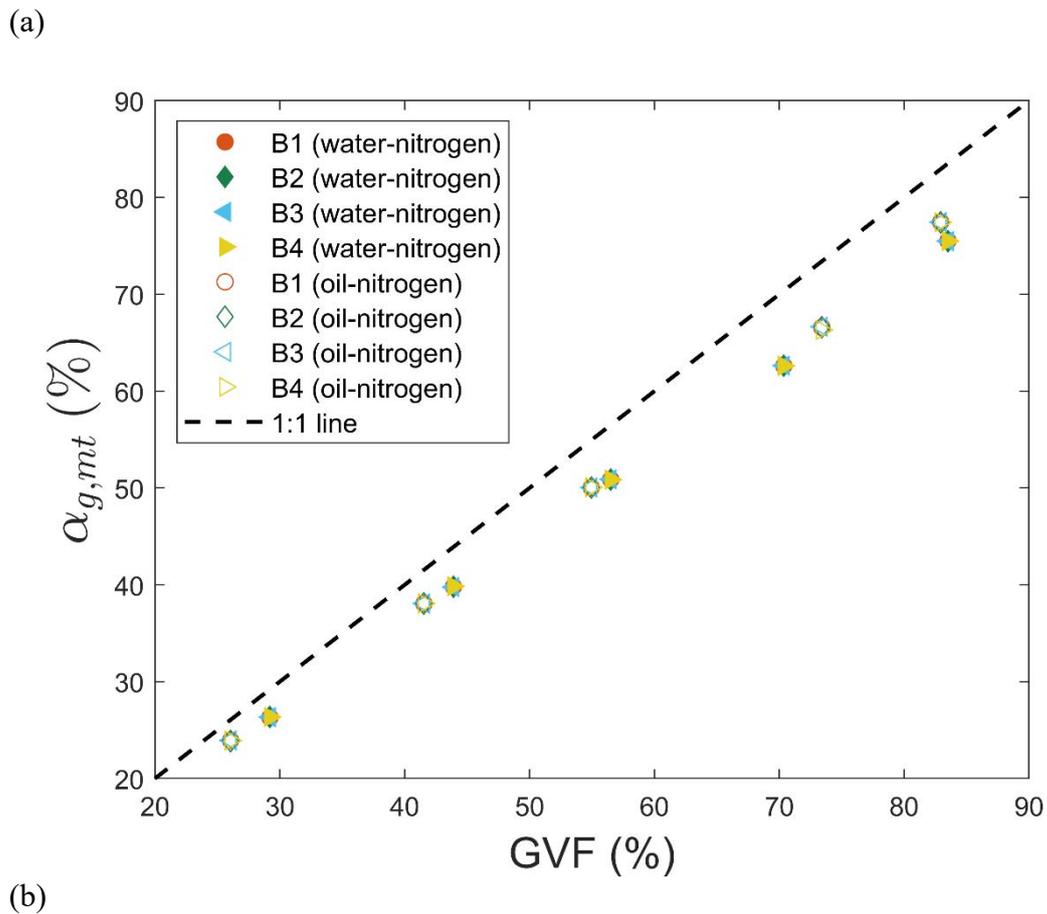

(b)

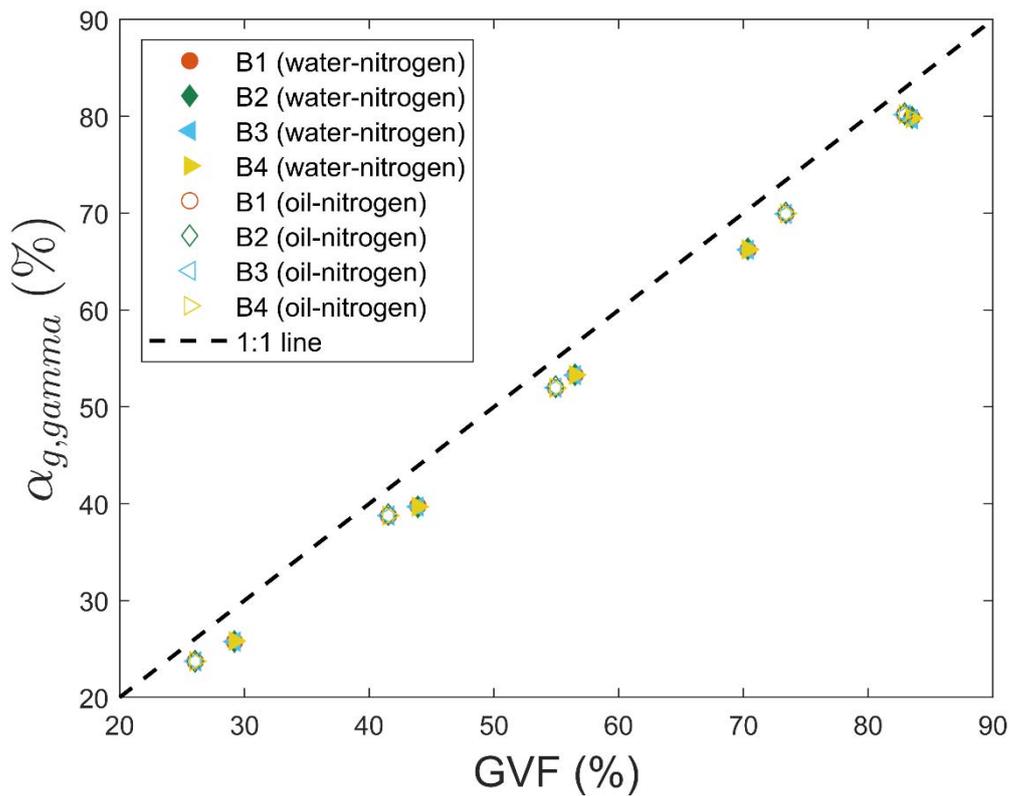

**Figure 13:** Variation of cross-sectional gas fraction at Venturi mid-throat $\alpha_{g,mt}$ (a) and gamma-beam-equivalent gas fraction $\alpha_{g,gamma}$ (b) with inlet gas volume fraction (GVF) for gas-liquid flow in flow domains B1, B2, B3, and B4 with increasing horizontal blind-tee depth.

It is observed that, unlike variation in VEL, which affects the phase distribution and $\alpha_{g,gamma}$, the difference in $\alpha_{g,mt}$ and $\alpha_{g,gamma}$ of the gas-liquid flow with different HBDs is insignificant. Variations in HBD have negligible effects on the phase distribution and the slip ratio $S$ downstream of the Venturi throat, where phase-fraction measurements are taken.

### 3.2.2 Venturi differential-pressure

Figure 14 shows the simulated differential-pressure $\Delta P$ between the Venturi mid-throat and the Venturi inlet for flow domains B1, B2, B3, and B4.

(a)

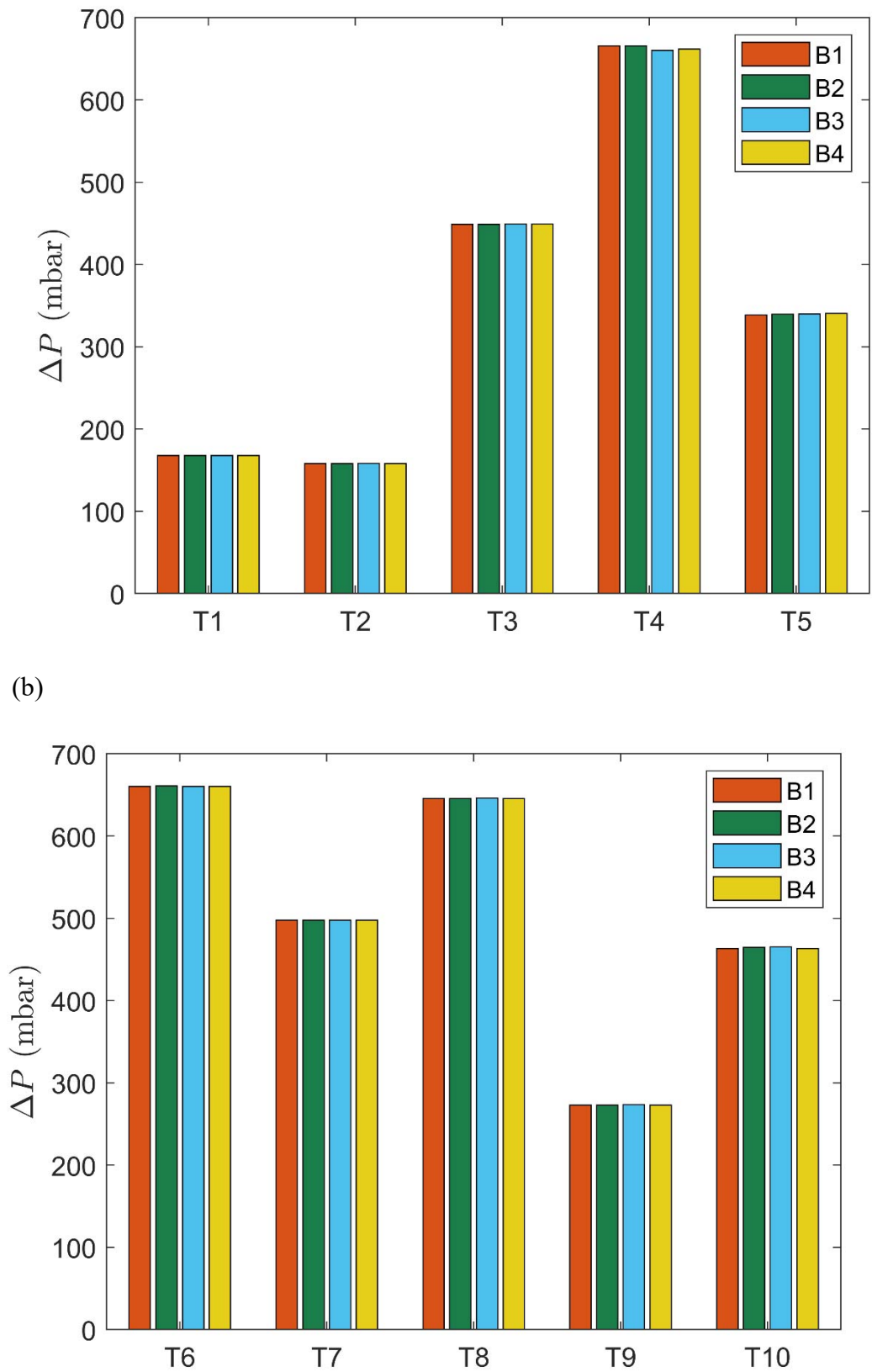

(b)

**Figure 14:** Simulated $\Delta P$ with increasing horizontal blind-tee depth in flow domains B1, B2, B3, and B4, for (a) water-nitrogen test points (T1 to T5); and (b) oil-nitrogen test points (T6 to T10).

A slight difference in the Venturi differential-pressure $\Delta P$ of the gas-liquid flow is observed for different HBDs. Hence, variations in HBD have negligible effects on the phase-fraction measured by the gamma-ray sensor and the Venturi differential-pressure measurement.

### 3.2.3  Local liquid-property in horizontal blind-tee

The lower region of a horizontal blind-tee spool upstream of a vertically mounted Venturi-based MPFM is a location for local liquid-property measurements due to the locally liquid-rich phase distribution. The magnitude of the local liquid tangential velocity is also important to ensure fluid exchange and measurement representativeness. To understand the effect of HBD on local liquid-property, the variations in local tangential velocity and liquid fraction of different HBDs near the end flanges of horizontal blind-tee are investigated. Since the measurement of local liquid properties at high-GVF test points with low liquid content is more challenging than at low-GVF test points, an example of HBD-effect analysis for high-GVF test points is shown in Figure 15.

(a)

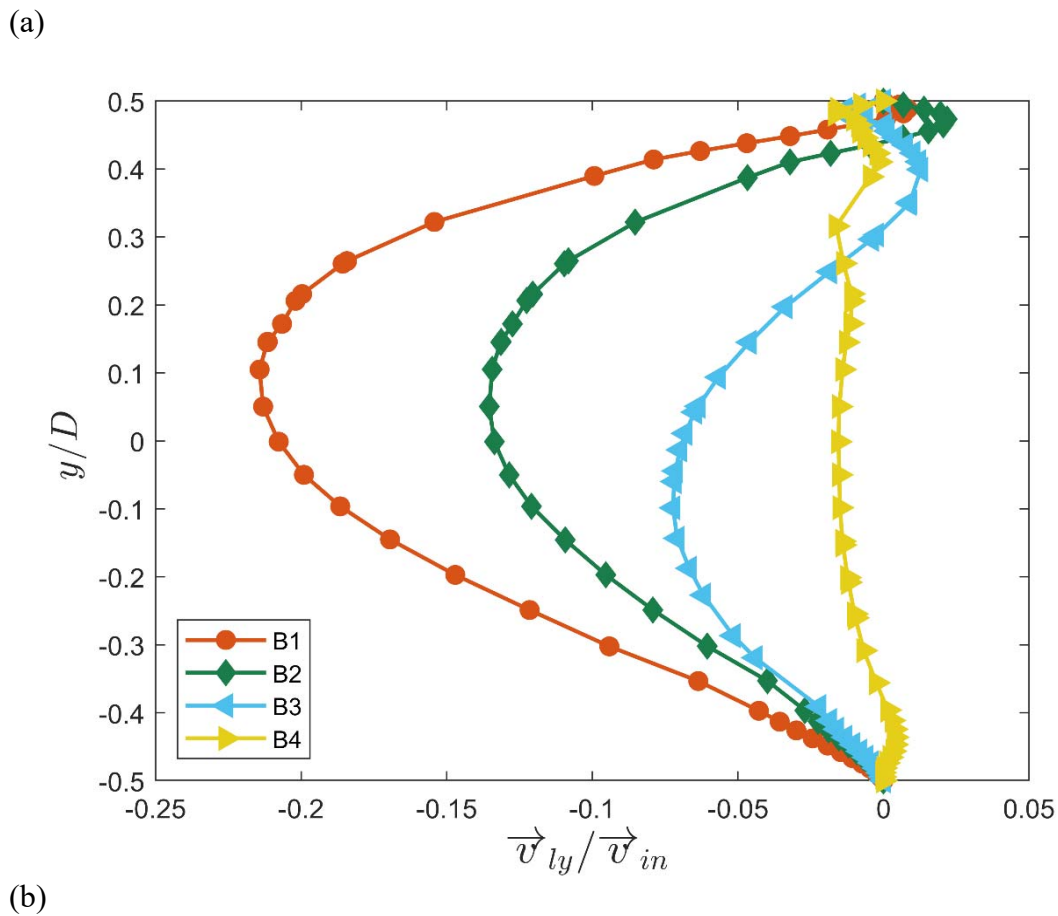

(b)

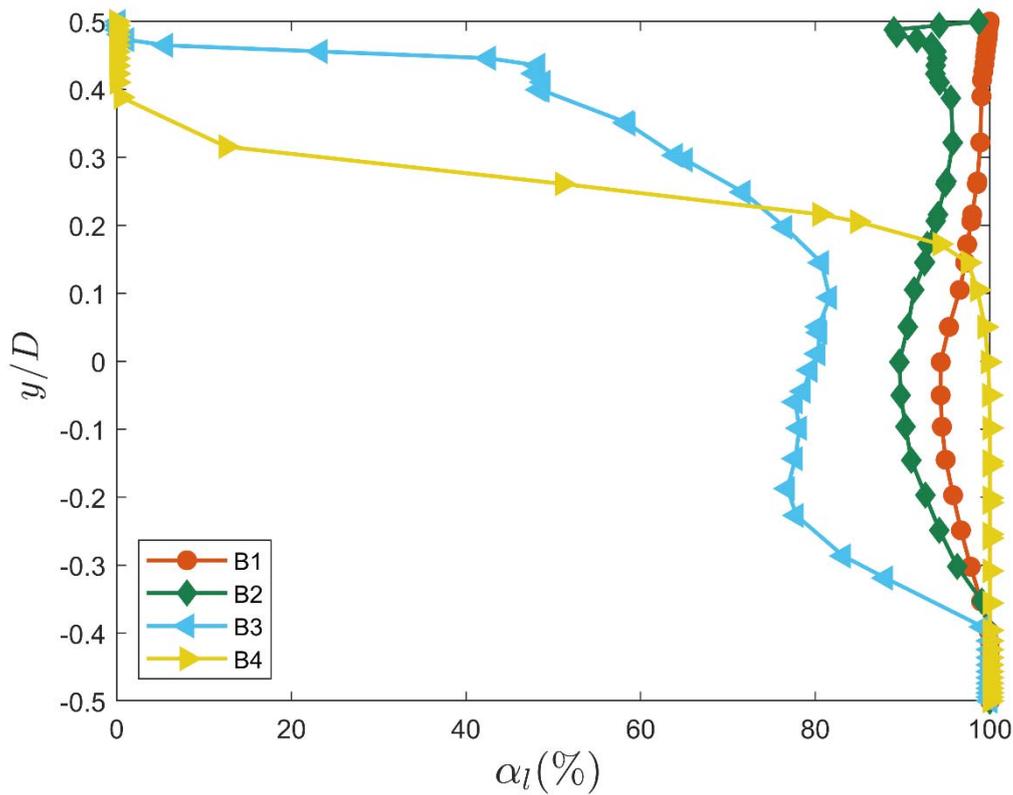

**Figure 15:** For increasing horizontal blind-tee depth (for flow domains B1, B2, B3, and B4), at 2 mm (sensing) depth away from end flange, variation of (a) normalized liquid tangential velocity against horizontal inlet velocity $\frac{v_{ly}}{|\bar{v}_{in}|}$; and (b) local liquid fraction $\alpha_l$, for oil-nitrogen flow with gas volume fraction 83%, homogeneous inlet velocity 6.07 m/s (T10).

The magnitude of the local liquid tangential velocity is observed to decrease with increasing HBD. For phase distribution, the extent of phase separation by gravity increases with increasing HBD, with an almost 100% liquid fraction $\alpha_l$ observed in the lower half of the horizontal blind-tee in B4; however, for the gas-liquid flow with relatively small HBD, the local phase distribution in the lower half of the horizontal blind-tee is still liquid-rich, with B1 and B2 having the minimal $\alpha_l$ above 90% and B3 having the minimal $\alpha_l$ above 70%. To get a deeper understanding of the flow field and fluid exchange in the horizontal blind-tee, a vector diagram of the liquid velocity in the horizontal blind-tee is shown in Figure 16.

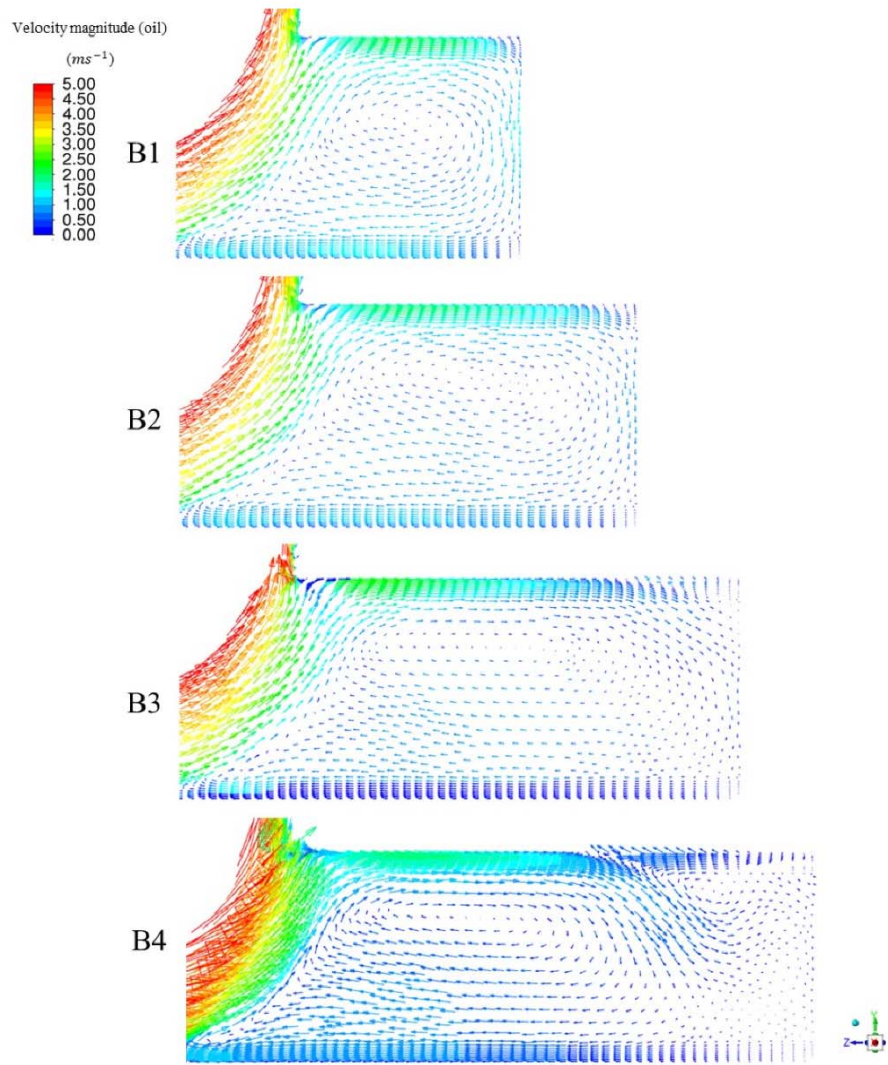

**Figure 16:** Vector of flow (oil) velocities inside horizontal blind-tee for oil-nitrogen flow with gas volume fraction
83%, homogeneous inlet velocity 6.07 m/s (T10) with increasing horizontal blind-tee depth in flow domains B1, B2,
B3, and B4.

Only one main circulation zone is observed in the horizontal blind-tee for B1 and B2 with

HBD $= 1.5D$ and $2D$, respectively. For B3 (with HBD $= 2.5D$), a small local vortex is observed in the upper right ($-z$) corner of the horizontal blind-tee; however, this is unlikely to affect the liquid properties measured in the lower half of the horizontal blind-tee. For B4 (HBD $= 3D$), in addition to the small local vortex at the top right corner, another local vortex is observed at the bottom half of the horizontal blind-tee, this may limit the fluid exchange in the local vortex with the fluid coming from the main pipe, which affects the representativeness or cause a time-lag for the real-time measurement of local liquid properties.

Hence, $\text{HBD} = 1.5D$ to $2D$ (B1 to B2) is suitable for liquid-property measurement as it fulfills the criteria of being liquid-rich in the lower half of the horizontal blind-tee and having sufficient local flow rate for fluid exchange. This conclusion applies to test points with a GVF range of $26\%$ to $83\%$ and an inlet homogeneous flow rate of $1.68\,\text{m/s}$ to $6.07\,\text{m/s}$. For flows with higher GVF, the liquid height $\frac{y}{D}$ in the horizontal blind-tee may be reduced. Hence, there may be a need to optimize sensor positioning to facilitate detection of liquid properties with low liquid content. For flows with lower flow velocities, the local vortices that limit fluid exchange may occur at shorter HBDs. Hence, for flow conditions beyond this range, further studies have to be performed to investigate the applicability of liquid-property measurements in horizontal blind-tee.

## 4   Experimental validation

Experimental validation is performed at the multiphase flow facility at the National University of Singapore. Rapid electrical capacitance measurements are used to determine the phase-fraction at both the Venturi inlet and mid-throat for oil-gas (non-conducting) flow by a pair of 8-electrode sensors connected to a commercial 16-channel AC-ECT (alternating current-based electrical capacitance tomography) system with high-frequency sinusoidal excitation and phase-sensitive demodulation (see Figure 17). The ECT system provides rapid, non-invasive phase-fraction measurement by measuring the effective capacitance between multiple pairs of sensing and detecting electrodes mounted circumferentially on the pipe-section (made of thick transparent dielectric material needed to visualize flow and withstand flow testing pressures), thus enabling a representative measurement of the fluid at the measuring pipe-cross-section(s) (bin Razali *et al*., 2021).

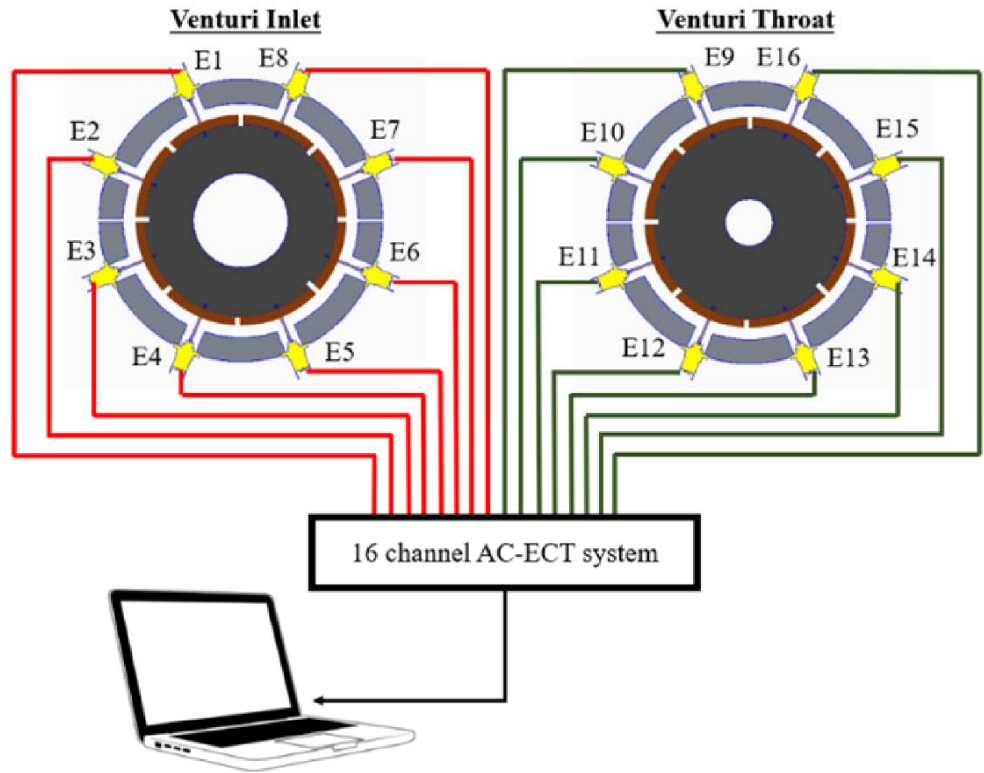

**Figure 17:** Sectional view of dual 8-electrode capacitance measurement sensor mounted at Venturi inlet and throat cross-section, and connection of electrodes to 16-channel AC-ECT (alternating current-based electrical capacitance tomography) system.

A multivariable transmitter is used to monitor line pressure $P$, $\Delta P$ and temperature $T$ of oil-air flow. Line pressure $P$ is ~ *3.5bar*, $T$ is ~ *30ºC*. The experimental validation is conducted for flow domains B1, C1, D1, and B4, with a GVF range of 15% to 95%, and a homogeneous inlet velocity range of 0.75 m/s to 9 m/s. The viscosity of the oil is ~ *16cP*.

### 4.1 Phase-fraction

The gas fraction at the Venturi inlet $\alpha_{g,vi(ECT)}$ and the Venturi mid-throat $\alpha_{g,mt(ECT)}$ measured by ECT for flow domains B1, C1, and D1 with increasing VEL is shown in Figure 18, while those for B1 and B4 with increasing HBD are shown in Figure 19.

(a)

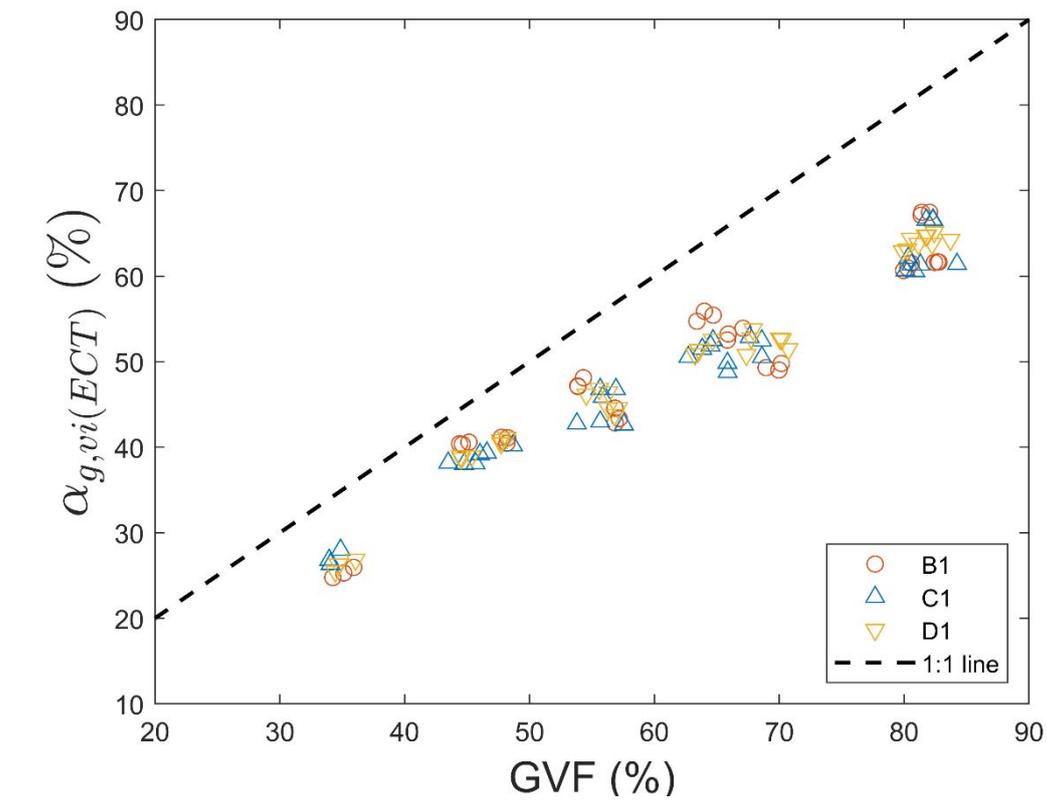

(b)

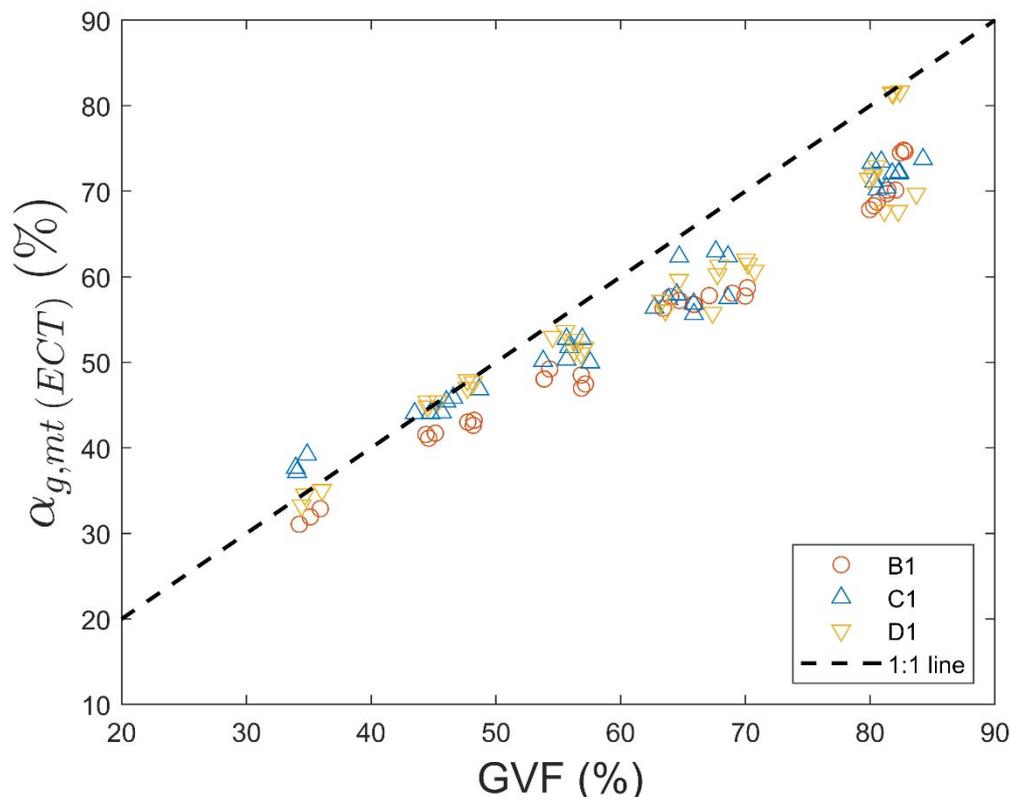

**Figure 18:** Electrical capacitance tomography (ECT) measured (oil-air flow) gas fraction at (a) Venturi inlet $\alpha_{g,vi(ECT)}$ and (b) Venturi mid-throat $\alpha_{g,mt(ECT)}$, for flow domains B1, C1, and D1 with increasing vertical entrance length.

(a)

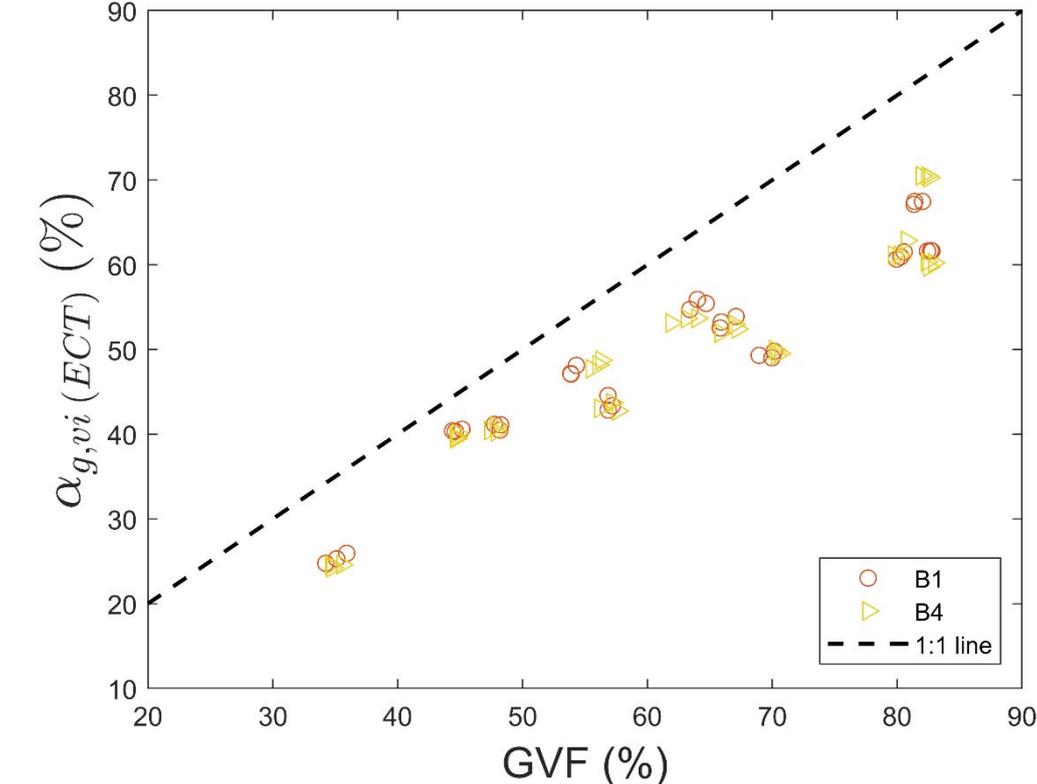

(b)

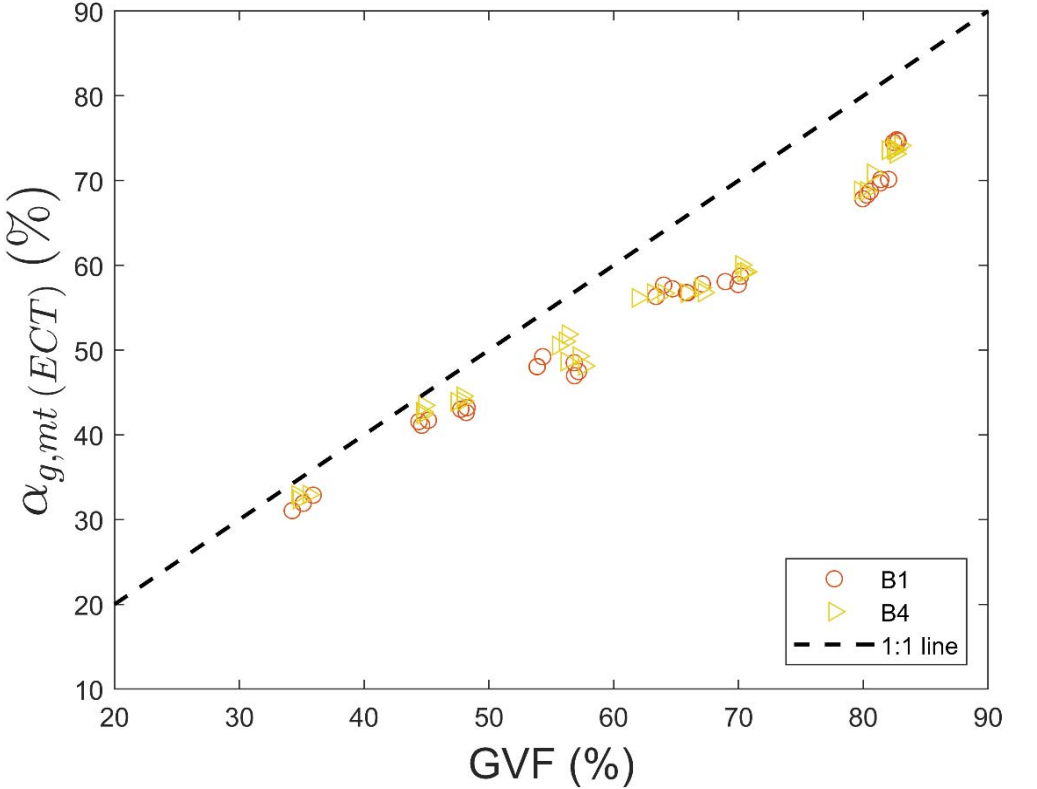

**Figure 19:** Electrical capacitance tomography (ECT) measured (oil-air flow) gas fraction at (a) Venturi inlet $\alpha_{g,vi(ECT)}$ and (b) Venturi mid-throat $\alpha_{g,mt(ECT)}$, for flow domains B1 and B4 with increasing horizontal blind-tee depth.

Figure 18 indicates that for $GVF < 40\%$ and $GVF > 80\%$, the difference in $\alpha_{g,vi(ECT)}$ is insignificant for oil-air flow in B1, C1, and D1. For $40\% \leq GVF \leq 80\%$, oil-air flow in B1 have greater $\alpha_{g,vi(ECT)}$ than that in C1 and D1 which have similar $\alpha_{g,vi(ECT)}$. The observation is similar to the simulation findings, where $\alpha_{vi}$ decreases with an increasing VEL in a range of 26% and 73% (oil-nitrogen flow), and the difference in $\alpha_{vi(ECT)}$ is insignificant for oil-nitrogen flow with different VELs for GVF of 83%.

At the Venturi mid-throat, for $GVF < 40\%$, $\alpha_{g,mt(ECT)}$ increases from B1 to C1. $\alpha_{g,mt(ECT)}$ at D1 is slightly smaller than C1. For $40\% \leq GVF \leq 80\%$, $\alpha_{g,mt(ECT)}$ increases from B1 to C1 while C1 and D1 share similar $\alpha_{g,mt(ECT)}$. The difference decreases with increasing VEL. For $GVF > 80\%$, the difference in $\alpha_{g,mt(ECT)}$ of oil-air flow with different VELs is negligible for most test points. The observation is similar to the simulation findings where $\alpha_{g,gamma}$ increases with an increasing VEL from B1 to C1 in a range of 26% and 83% (oil-nitrogen flow), and the difference in $\alpha_{g,gamma}$ is insignificant between C1 and D1. The reason why the observation of $\alpha_{g,mt(ECT)}$ is more similar to $\alpha_{g,gamma}$ than $\alpha_{g,mt}$ is that, even with circumferentially mounted electrodes, soft-electromagnetic-field ECT measurement does not cover the entire pipe-cross-section and is still subjected to the effect of phase distribution.

From Figure 19, it is observed that the oil-air flow in B1 and B4 with increasing HBD have similar $\alpha_{g,vi(ECT)}$ and $\alpha_{g,mt(ECT)}$. This is consistent with the simulation findings that variations in HBD have a negligible effect on gas fraction measurements at the Venturi throat.

### 4.2 Venturi differential-pressure

Figure 20 shows that the relative difference in $\Delta P$ between flows in flow domains B1 and D1 is smaller than $\pm 3\%$ for most of the test points, except for one water-air point with a low $\Delta P$ of 121.7 mbar (B1) having a relative difference of ~ 4% with D1. Hence, variation in VEL has no significant impact on $\Delta P$ measurements of VEL equal to or greater than $6D$. The discussion in Section 4.1.1 indicates that variation in HBD does not affect the cross-sectional gas fraction at both the Venturi inlet and the Venturi mid-throat. From the mass flow conservation and the Bernoulli equation (Equation 1), the differential-pressure is also not expected to be markedly affected by variation in HBD.

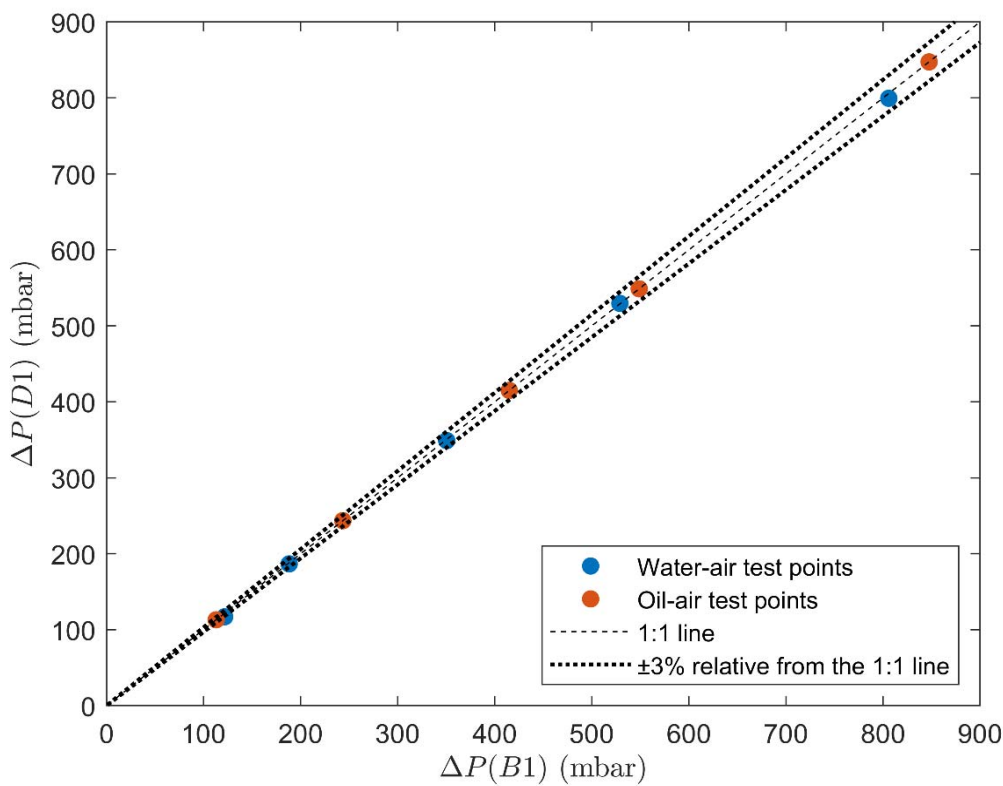

**Figure 20:** Measured $\Delta P$ in flow domain D1 versus measured $\Delta P$ in flow domain B1, National University of Singapore flow-loop test.

## 5 Conclusions

In this study, the effect of variation in VEL and HBD on multiphase flow measurement of phase-fraction and differential-pressure in the vertical Venturi, as well as the effect of liquid properties at the ends of horizontal blind-tee, are investigated by using CFD or experiments. It is found that even though the CFD-simulated cross-sectional phase-fraction at the Venturi mid-throat $\alpha_{g,mt}$ are similar for gas-liquid flow with different VELs, the phase-fraction measured by gamma or ECT sensors increases with increasing VEL in the range of GVF 26% to GVF 83% for gamma-beam-equivalent phase-fraction $\alpha_{g,gamma}$ (simulated), and GVF 40% to GVF 80% for phase-fraction measured by ECT $\alpha_{g,mt(ECT)}$ (experimental). This is because the chordal measurements by the gamma-beam sensor and ECT measurements have a depedence on the phase distribution of the Venturi throat, which has an increasing axisymmetry with increasing VEL. From the simulation results, the differential-pressure increases slightly with increasing VEL, because a greater amount of work done by pressure is required to accelerate liquid in the flow with a longer VEL where gas-liquid slip is larger (or liquid velocity is lower) at the Venturi inlet to the Venturi throat; however, the difference is insignificant to be reflected in the experiments with VEL above $6D$. The discharge coefficient also increases with increasing VEL, probably because gas-liquid flow with a shorter VEL has a thicker near-wall liquid layer, and is subjected to a greater frictional loss. Unlike variations in VEL, variations in HBD have negligible effects on phase-fraction and differential-pressure measurements; however, an excessive HBD ($3D$) may cause localized vortices and low liquid tangential velocity at the end of the horizontal blind-tee. This may limit local fluid exchange with the main pipe, leading to inaccurate measurements of liquid properties.

The main contributions of this work include: (1) The main effects of horizontal blind-tee design parameters, VEL and HBD, on multiphase flow measurements, including phase-fraction, differential-pressure, and local liquid properties in a vertically mounted Venturi is identified for a range of flow conditions; (2) A suitable horizontal blind-tee design with VEL of $6D$ and HBD of $1.5D$ to $2D$ is recommended for multiphase measurements in the consideration of flow model accuracy, pipe-section cost savings and, hence, carbon footprint reduction.

## Acknowledgements

The authors are grateful to Chan W.L., Basman E., and Hammond P. for constructive comments and suggestions. This work was supported by Singapore Economic Development

## References

Acharya, T., & Casimiro, L. (2020). Evaluation of flow characteristics in an onshore horizontal separator using computational fluid dynamics. *Journal of Ocean Engineering and Science*, **5**(3), 261–268.

Baker, O. (1953). Design of pipelines for the simultaneous flow of oil and gas. *Proceedings of the Fall Meeting of the Petroleum Branch of AIME*, 323-G.

Burns, A.D., Frank, T., Hamill, I., & Shi, J.-M. (2004). The Favre averaged drag model for turbulent dispersion in Eulerian multi-phase flows. *Proceedings of the 5th International Conference on Multiphase Flow*, **392**, 1–17.

Cokljat, D., Slack, M., Vasquez, S.A., Bakker, A., & Montante, G. (2006). Reynolds-stress model for Eulerian multiphase. *Progress in Computational Fluid Dynamics*, **6**(1-3), 168–178.

Coughtrie, A.R., Borman, D.J., & Sleigh, P.A. (2013). Effects of turbulence modelling on prediction of flow characteristics in a bench-scale anaerobic gas-lift digester. *Bioresource Technology*, **138**, 297–306.

Fiore, M., Xie, C.-G., & Jolivet, G. (2019). Wider salinity range and lower water detection limit for multiphase flowmeters. *Proceedings of the SPE/IATMI Asia Pacific Oil & Gas Conference and Exhibition*.

Frank, T., Zwart, P.J., Krepper, E., Prasser, H.-M., & Lucas, D. (2008). Validation of CFD models for mono- and polydisperse air-water two-phase flows in pipes. *Nuclear Engineering and Design*, **238**(3), 647–659.

Han, F.H., Liu, Y.X., Ong, M.C., Yin, G., Li, W.H., & Wang, Z. (2022). CFD investigation of blind-tee effects on flow mixing mechanism in subsea pipelines. *Engineering Applications of Computational Fluid Mechanics*, **16**(1), 1395–1419.

Han, F.H., Ong, M.C., Xing, Y.H., & Li, W.H. (2020). Three-dimensional numerical investigation of laminar flow in blind-tee pipes. *Ocean Engineering*, **217**, 107962.

Hjertaker, B.T., Tjugum, S.-A., Hallanger, A., & Maad, R. (2018). Characterization of multiphase flow blind-T mixing using high speed gamma-ray tomometry. *Flow Measurement and Instrumentation*, **62**, 205–212.

Huang, S., Xie, C., Lenn, C., Yang, W., & Wu, Z. (2013). Issues of a combination of ultrasonic Doppler velocity measurement with a Venturi for multiphase flow metering. *Proceedings of the 18th Middle East Oil & Gas Show and Conference*, **1**, 2101–2109.

Laleh, A.P., Svrcek, W.Y., & Monnery, W.D. (2011). Computational fluid dynamics simulation of pilot-plant-scale two-phase separators. *Chemical Engineering & Technology*, **34**(2), 296–306.

López, J., Pineda, H., Bello, D., & Ratkovich, N. (2016). Study of liquid-gas two-phase flow in horizontal pipes using high speed filming and computational fluid dynamics. *Experimental Thermal and Fluid Science*, **76**, 126–134.

Menter, F.R. (1994). Two-equation eddy-viscosity turbulence models for engineering applications. *AIAA Journal*, **32**(8), 1598–1605.

Milovan, P., & Stephen, F. (2004). The advantage of polyhedral meshes, CD-adapco.

Pinguet, B., Smith, M.T., Vagen, N., Alendal, G.M., Rustad, R., & Xie, C.-G. (2014). An innovative liquid detection sensor for wet gas subsea business to improve gas-condensate flow rate measurement and flow assurance issue. *Proceedings of the Offshore Technology*

*Conference Asia*, **1**, 25054.

bin Razali, M.A., Xie, C.-G., & Loh, W.L. (2021). Experimental investigation of gas-liquid flow in a vertical Venturi installed downstream of a horizontal blind tee flow conditioner and the flow regime transition. *Flow Measurement and Instrumentation*, **80**, 101961.

Shang, Z., Lou, J., & Li, H.Y. (2015). A new multidimensional drift flux mixture model for gas-liquid droplet two-phase flow. *International Journal of Computational Methods*, **12**(4), 1540001.

Shu, J.-J. (2003a). A finite element model and electronic analogue of pipeline pressure transients with frequency-dependent friction. *Journal of Fluids Engineering-Transactions of the ASME*, **125**(1), 194–199.

Shu, J.-J. (2003b). Modelling vaporous cavitation on fluid transients. *International Journal of Pressure Vessels and Piping*, **80**(3), 187–195.

Shu, J.-J., Burrows, C.R., & Edge, K.A. (1997). Pressure pulsations in reciprocating pump piping systems Part 1: Modelling. *Proceedings of the Institution of Mechanical Engineers Part I-Journal of Systems and Control Engineering*, **211**(3), 229–237.

Shu, J.-J., Teo, J.B.M., & Chan, W.K. (2016). A new model for fluid velocity slip on a solid surface. *Soft Matter*, **12**(40), 8388–8397.

Shu, J.-J., Teo, J.B.M., & Chan, W.K. (2017). Fluid velocity slip and temperature jump at a solid surface. *Applied Mechanics Reviews*, **69**(2), 020801.

Shu, J.-J., Teo, J.B.M., & Chan, W.K. (2018). Slip of fluid molecules on solid surfaces by surface diffusion. *PLoS One*, **13**(10), e0205443.

Shu, J.-J., & Wilks, G. (1995). An accurate numerical method for systems of differentio-integral equations associated with multiphase flow. *Computers & Fluids*, **24**(6), 625–652.

Tomiyama, A., Kataoka, I., Zun, I., & Sakaguchi, T. (1998). Drag coefficients of single bubbles under normal and micro gravity conditions. *JSME International Journal Series B-Fluids and Thermal Engineering*, **41**(2), 472–479.

Tomiyama, A., Tamai, H., Zun, I., & Hosokawa, S. (2002). Transverse migration of single bubbles in simple shear flows. *Chemical Engineering Science*, **57**(11), 1849–1858.

Wilcox, D.C. (1997). Turbulence Modeling for CFD, D C W Industries, second edition.

Yamoah, S., Martínez-Cuenca, R., Monrós, G., Chiva, S., & Macián-Juan, R. (2015). Numerical investigation of models for drag, lift, wall lubrication and turbulent dispersion forces for the simulation of gas-liquid two-phase flow. *Chemical Engineering Research & Design*, **98**, 17–35.

Zeghloul, A., Azzi, A., Saidj, F., Azzopardi, B.J., & Hewakandamby, B. (2015). Interrogating the effect of an orifice on the upward two-phase gas-liquid flow behavior. *International Journal of Multiphase Flow*, **74**, 96–105.

Zhan, M.K., Xie, C.-G., & Shu, J.-J. (2022). Microwave probe sensing location for Venturi-based real-time multiphase flowmeter. *Journal of Petroleum Science and Engineering*, **218**, 111027.

Zhang, W.W., Yu, Z.Y., & Li, Y.J. (2019). Application of a non-uniform bubble model in a multiphase rotodynamic pump. *Journal of Petroleum Science and Engineering*, **173**, 1316–1322.